\documentclass[acmsmall,natbib=true,screen=true]{acmart}

\setcopyright{rightsretained}
\copyrightyear{2024}
\acmDOI{}

\acmJournal{TOIS}
\acmVolume{*}
\acmNumber{*}
\acmArticle{1}
\acmYear{2024}
\acmMonth{6}

\usepackage{color}
\usepackage{bbm}
\usepackage{multirow}
\usepackage[inline]{enumitem}
\usepackage{graphicx}
\usepackage{subcaption}
\usepackage{sistyle}
\usepackage[ruled]{algorithm2e}
\usepackage{booktabs}
\usepackage[skip=1pt]{caption}
\SIthousandsep{,}
\usepackage{makecell}
\usepackage{amsmath}
\usepackage[english]{babel}
\usepackage{hyperref}
\usepackage{array} 
\usepackage{graphicx}
\usepackage{algorithmic}
\usepackage{amsmath}
\usepackage{wrapfig}
\usepackage[figuresright]{rotating}

\definecolor{c1}{RGB}{255,255,157}
\definecolor{c2}{RGB}{190, 235, 159}
\definecolor{c3}{RGB}{94,180,210}

\usepackage{tikz}
\usetikzlibrary{trees}
\usetikzlibrary{positioning}
\usepackage{tikz-qtree}

\tikzset{
    bone/.style={text width=2cm, align=center},
    line/.style={thick, -latex},
}

\usepackage{acronym}

\usepgflibrary{arrows.meta}
\usetikzlibrary{positioning,calc,shapes}
\newtheorem{definition}{Definition}[section]

\newcommand{\heading}[1]{\vspace*{1mm}\noindent\textbf{#1.}}

\newcommand\blfootnote[1]{%
  \begingroup
  \renewcommand\thefootnote{}\footnote{#1}%
  \addtocounter{footnote}{-1}%
  \endgroup
}
\AtBeginDocument{%
  \providecommand\BibTeX{{%
    \normalfont B\kern-0.5em{\scshape i\kern-0.25em b}\kern-0.8em\TeX}}}

\definecolor{inkblue}{RGB}{0, 79, 152}

\acrodef{CV}{computer vision}
\acrodef{DRM}{dense retrieval model}
\acrodef{GR}{generative retrieval}
\acrodef{IID}{independent and identically distributed}
\acrodef{IR}{information retrieval}
\acrodef{LLM}{large language model}
\acrodef{NLP}{natural language processing}
\acrodef{NRM}{neural ranking model}
\acrodef{OOD}{out-of-distribution}
\acrodef{SEO}{search engine optimization}

\definecolor{bgblue}{HTML}{26BDE2}

\begin{document}

\title{Robust Neural Information Retrieval: An Adversarial and Out-of-distribution Perspective}

\author{Yu-An Liu}
\orcid{0000-0002-9125-5097}
\email{liuyuan21b@ict.ac.cn}

\author{Ruqing Zhang}
\orcid{0000-0003-4294-2541}
\email{zhangruqing@ict.ac.cn}  

\author{Jiafeng Guo}  
      \authornote{Jiafeng Guo is the corresponding author.}
\orcid{0000-0002-9509-8674}
\email{guojiafeng@ict.ac.cn}    

\affiliation{
\institution{CAS Key Lab of Network Data Science and Technology, Institute of Computing Technology, Chinese Academy of Sciences; University of Chinese Academy of Sciences}
  \streetaddress{No. 6 Kexueyuan South Road, Haidian District}
  \city{Beijing}
  \country{China}
  \postcode{100190}
}

\author{Maarten de Rijke}
\orcid{0000-0002-1086-0202}
\affiliation{
 \institution{University of Amsterdam}
 \city{Amsterdam}
 \country{The Netherlands}
}
\email{m.derijke@uva.nl}

\author{Yixing Fan}
\orcid{0000-0003-4317-2702}
\email{fanyixing@ict.ac.cn}

\author{Xueqi Cheng}
\orcid{0000-0002-5201-8195}
\email{cxq@ict.ac.cn}
\affiliation{
\institution{CAS Key Lab of Network Data Science and Technology, Institute of Computing Technology, Chinese Academy of Sciences; University of Chinese Academy of Sciences}
 \city{Beijing}
 \country{China}
}

\renewcommand{\shortauthors}{Liu et al.}

\begin{abstract}
Recent advances in neural information retrieval models have significantly enhanced these models' effectiveness across various information retrieval tasks. 
The robustness of these models, which is essential for ensuring their reliability in practice, has also garnered significant attention. 
With a wide array of research on robust IR being published, we believe it is the opportune moment to consolidate the current status, glean insights from existing methodologies, and lay the groundwork for future development. 
We view the robustness of information retrieval to be a multifaceted concept, emphasizing its necessity against adversarial attacks, out-of-distribution scenarios, and performance variance. 
With a focus on adversarial and \acl{OOD} robustness, we dissect robustness solutions for \aclp{DRM} and \aclp{NRM}, respectively, recognizing them as pivotal components of the neural \acl{IR} pipeline. 
We provide an in-depth discussion of methods, datasets, and evaluation metrics, shedding light on challenges and future directions in the era of large language models. 
As side-products of this survey, we release three additional resources:
\begin{enumerate*}[label=(\roman*)]
\item a curated list of publications related to robust IR,\footnote{\url{https://github.com/Davion-Liu/Awesome-Robustness-in-Information-Retrieval}\label{fn:reading-list}} 
\item a tutorial based on this survey,\footnote{\url{https://sigir24-robust-information-retrieval.github.io}\label{fn:tutorial}} and
\item a heterogeneous benchmark for robust \acl{IR}, BestIR, that collects all known datasets for evaluation \acl{IR} systems for robustness.\footnote{\url{https://github.com/Davion-Liu/BestIR}\label{fn:benchmark}}
\end{enumerate*}
We hope that this study provides useful clues for future research on the robustness of IR models and helps to develop trustworthy search engines.
\end{abstract}



\maketitle

\acresetall

\section{Introduction}
Recently, with advances in deep learning, neural \ac{IR} models have witnessed significant progress \cite{guo2020deep,guo2022semantic}. 
With the development of training methodologies such as pre-training \cite{gao2022unsupervised, ma2022contrastive} and fine-tuning \cite{Qu2021RocketQAAO,Zhan2021OptimizingDR,Khattab2020ColBERTEA}, neural \ac{IR} models have demonstrated remarkable effectiveness in learning query-document relevance patterns.
In deployment of neural \ac{IR} models, an aspect equally essential as their effectiveness is their robustness.  
A good \ac{IR} system must not only exhibit high effectiveness under normal conditions but also demonstrate robustness in the face of abnormal conditions.

\paragraph{Why is robustness important in \ac{IR}\@?}
The natural openness of \ac{IR} systems makes them vulnerable to intrusion, and the consequences can be severe.
For example: 
\begin{enumerate*}[label=(\roman*)]
    \item search engines are vulnerable to black hat \ac{SEO} attacks,\footnote{\url{https://www.bleepingcomputer.com/news/security/15-000-sites-hacked-for-massive-google-seo-poisoning-campaign/}} necessitating significant efforts to curb these infringements.\footnote{\url{https://www.bbc.com/news/technology-28687513},  \url{https://developers.google.com/search/docs/essentials}} and
    \item search engines are confronted with large amounts of unseen data on a daily basis. The working algorithm needs to be improved constantly to ensure that search effectiveness is maintained.\footnote{\url{https://developers.google.com/search/news}}
\end{enumerate*}

Recently, academic researchers have begun to investigate the robustness of \ac{IR} systems \cite{wu2022neural,thakur2beir,cohen2018cross,liu2022order,liu2023topic,chen2023towards}. 
As neural networks gain increasing popularity in IR, many studies have found that neural IR systems inherit a wide variety of problematic robustness issues from deep neural networks.
In response, the field of robust neural \ac{IR} is garnering increasing attention, as evidenced by the growing number of papers published annually, as depicted in Figure~\ref{fig:publications}.\footnote{See Appendix~\ref{appendix:A} for a description of the protocol we followed to select the sources aggregated in Figure~\ref{fig:publications} and surveyed in this paper.}
The robustness issues are differently represented in real \ac{IR} scenarios and raise concerns about deploying neural \ac{IR} systems into the real world.
Therefore, the study of robust neural IR is crucial for building reliable IR systems.

\paragraph{How to defined robustness in IR\@?}
User attention mainly focuses on the Top-$K$ results and achieves with higher rankings \cite{niu2012top}.
Based on this, we argue that robustness in \ac{IR} refers to the consistent performance and resilience on the Top-$K$ results of an \ac{IR} system when faced with a variety of unexpected scenarios.
Robustness is not a simple concept; it encompasses multiple dimensions, as illustrated by research within the machine learning (ML) community \cite{shafique2020robust,zhang2020machine}. 
In \ac{IR}, we identify several facets of robustness:  
\begin{enumerate}[label=(\arabic*)]
    \item \textit{\Acfi{IID} robustness} emphasizes the worst-case performance across different individual queries under the \ac{IID} data assumption \cite{wu2022neural};
    \item \Acfi{OOD} robustness refers to the generalizability of an IR model on unseen queries and documents from different distributions of the training dataset \cite{thakur2beir};  and
    \item \textit{Adversarial robustness} refers to the ability of the IR model to defend against malicious adversarial attacks aimed at manipulating rankings \cite{wu2022neural}.
\end{enumerate}
\begin{figure}[t]
    \centering
    \includegraphics[width=\linewidth]{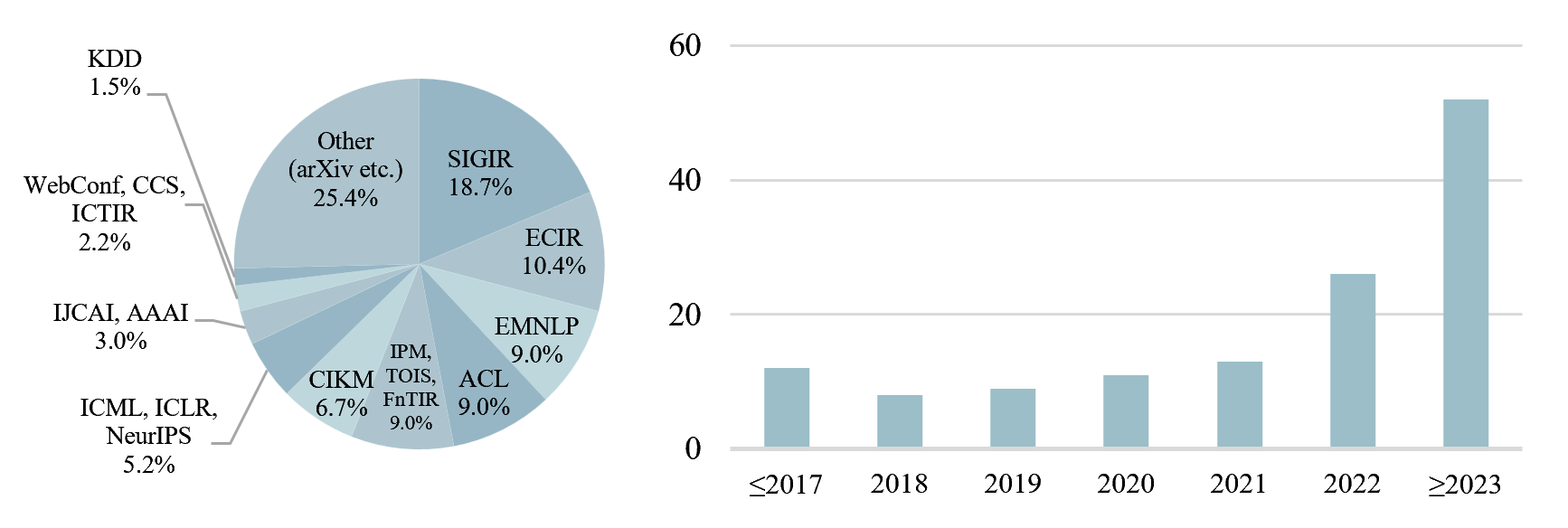}
    \caption{Statistics of publications related to robust neural information retrieval and covered in this survey.
    ``Other” includes arXiv (mostly), TREC, ICDM, NAACL, and TACL.}
    \label{fig:publications}
\end{figure}

\noindent%
In this survey, our focus is on adversarial robustness and \ac{OOD} robustness, which have garnered significant attention:
\begin{enumerate*}[label=(\roman*)]
    \item for adversarial robustness, studies primarily approach the topic from two angles, i.e., adversarial attack and defense, to enhance the robustness of \ac{IR} models, and
    \item for \ac{OOD} robustness, the emphasis is on improving the generalizability of \ac{IR} models to both unseen documents and unseen queries.
\end{enumerate*}
To study the above two aspects, we zoom in on two key components of the neural IR framework: first-stage retrieval and the subsequent ranking stage. 
We focus on \emph{\aclp{DRM}} (\acsp{DRM}) and \emph{\aclp{NRM}} (\acsp{NRM}) to further explore the aforementioned research perspectives.

\paragraph{Relation to other surveys}
There are several surveys on robustness in the fields of \ac{NLP} \citep{wang2022measure, Wang2023TowardsAR} and \ac{CV} \citep{drenkow2021systematic,akhtar2021advances}.
However, the field of IR presents its own unique characteristics: 
\begin{enumerate*}[label=(\roman*)]
    \item unlike \ac{NLP}, which often focuses on individual examples, \ac{IR} involves ranking a collection of documents, highlighting the need for robustness across the ranked lists, and 
    \item different from continuous image data in \ac{CV}, \ac{IR} deals with robustness concerning discrete text documents.
\end{enumerate*}
Consequently, the studies explored in these surveys are not directly transferrable as references within the IR field. 

Surveys specific to the \ac{IR} domain tend to concentrate on areas like pre-training \cite{zhao2024dense,fan2022pre}, ranking models \cite{guo2020deep}, initial retrieval stages \cite{guo2022semantic}, and the explainability of \ac{IR} systems \cite{anand2023explainable}.
Thus, there is a noticeable gap in the literature: a dedicated survey that consolidates and introduces research pertaining to robustness in IR is absent.

On top of the work discussed above, this survey adds the following:
\begin{enumerate*}[label=(\roman*)]
    \item our work complements robustness surveys from other fields, taking into account the distinctive characteristics of different domains, and
    \item our survey complements \ac{IR} surveys by not only prioritizing effectiveness but also emphasizing robustness. 
\end{enumerate*}
To complement this survey we also release the following resources alongside it: 
\begin{enumerate*}[label=(\roman*)]
    \item a curated list of publications related to robust \ac{IR},
    \item a tutorial on robust \ac{IR}, and
    \item a benchmark that collects datasets for assessing the robustness of \ac{IR} systems.
\end{enumerate*}
See footnotes~\ref{fn:reading-list}, \ref{fn:tutorial}, and \ref{fn:benchmark}, respectively.

\paragraph{Contributions of this survey}
We present a compilation of research in IR robustness, summarizing and organizing the content to aid understanding. 
By categorizing each subfield and delineating its core concepts, we aim to facilitate a deeper understanding of this field and promote the long-term development of robust neural IR. 
In summary, this paper's contributions are as follows:
\begin{itemize}
    \item A comprehensive overview and categorization: we define robustness in \ac{IR} by summarizing the literature and further dividing it into distinct categories; we provide a curated list of publications to support this aspect of the survey.
    \item A detailed discussion of methodologies and datasets: we offer a detailed discussion of methodologies, datasets, and evaluation metrics pertinent to each aspect of robustness.
    We support this part of the survey by providing a tutorial and a benchmark, BestIR, which integrates the datasets mentioned in this survey to facilitate follow-up work.
    \item Identification of open issues and future trends: we highlight challenges and potential future trends, particularly in the age of \acp{LLM}.
\end{itemize}

\begin{figure}[t]
    \centering
    \includegraphics[width=\linewidth]{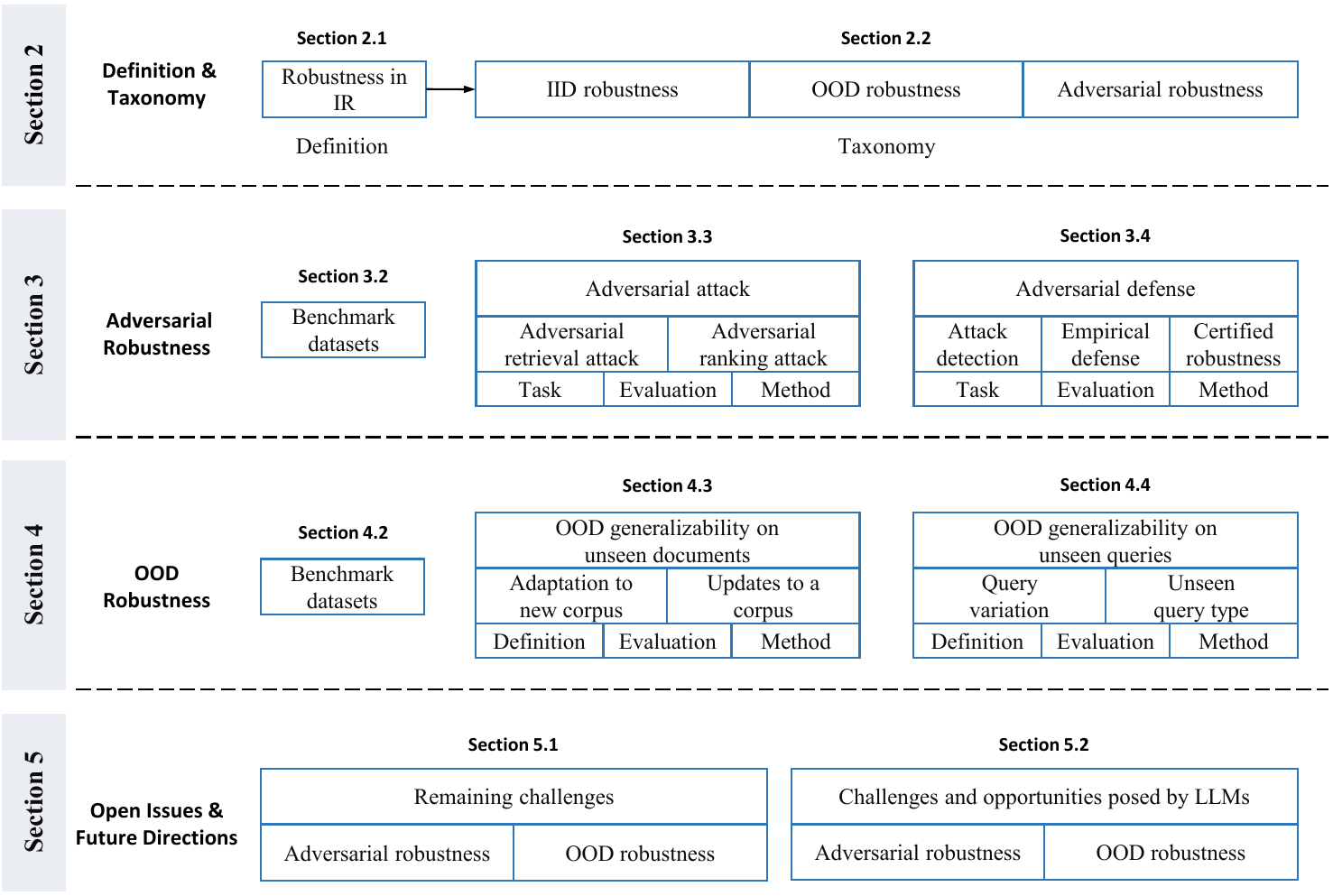}
    \caption{Overview of the survey. Section \ref{section Future} is only partially listed here because of space limitations.}
    \label{fig:overview}
\end{figure}

\paragraph{Organization} 
Figure~\ref{fig:overview} depicts the organization of our survey.
In Section~\ref{section 2}, we introduce the \ac{IR} task, and give a definition and taxonomy of robustness in \ac{IR}.
In Section~\ref{Sec: Adversarial Robustness}, we examine adversarial attack and defense tasks, alongside their respective datasets, evaluation criteria, and state-of-the-art methodologies. 
In Section~\ref{Sec: Out-of-distribution Robustness}, we show two key scenarios for OOD robustness, i.e., \ac{OOD} generalizability on unseen documents and \ac{OOD} generalizability to unseen queries, and present specific datasets, evaluation metrics, and methods for solving these scenarios. 
In Section~\ref{section Future}, we describe remaining challenges and emerging opportunities for robustness of \ac{IR} in the era of \acp{LLM}. 
Finally, Section~\ref{section Conclusion} summarizes the survey and offers concluding remarks. 

\section{Definition and Taxonomy} \label{section 2}
In this section, we provide a formal definition of robustness in the context of IR
and outline the taxonomy pertinent to this domain.

\paragraph{IR task}
To provide a clear understanding, we first formalize the IR task.
Suppose that $R = \{r_1, r_2, \ldots, r_l\}$ is the set of relevance levels, where $l$ denotes the number of levels.
A total order exists among the relevance labels such that $r_l \succ r_{l-1} \succ \cdots \succ r_1$, where $\succ$ denotes the order relation.
The minimum value of the relevance label is 0, which usually implies no relevance.
Suppose that $Q = \{q_1, q_2, \ldots, q_m\}$ is the set of queries in the training dataset. 
Each query $q_i$ is associated with a list of documents $D_i=\{d_{i,1}, d_{i,2}, \ldots, d_{i,N} \}$ and a list of relevance labels $Y_i = \{y_{i,1}, y_{i,2}, \ldots ,y_{i,N}\}$, where $y_{i,j} \in R$ denotes the label of document $d_{i,j}$ and $N$ is the document list size. 
Then we obtain the training dataset $\mathcal{D}_\mathrm{train} = \{(q_i, D_i, Y_i)\}^{m}_{i=1}$.

We use $f$ to denote the IR model; it predicts the relevance score $f(q,d)$ based on a given query $q$ and document $d$. 
The IR model $f$ is derived by learning from the following objective:
\begin{equation}
        \theta^* = \arg \min_\theta  \mathbb{E}_{(q, D, Y) \sim \mathcal{D}_\mathrm{train}} 
        \ \mathcal{L}\left(f(q,d),Y\right),
\label{equ: learning_objective}
\end{equation} 
where $\theta$ are the parameters of the IR model $f$, and $\mathcal{L}$ is a ranking loss function.

We employ the ranking function $f$ to generate a score for any query-document pair $(q,d)$, reflecting the relevance degree of $d$ given $q$.
This allows us to produce a rank $\pi_f\left(q, d\right)$ for each $d \in D$ according to the predicted score.
In general, the ranking function $f$ can be further abstracted by the following unified formulation:
\begin{equation}
	f(q,d)=g\left(\psi(q),\phi(d),\eta(q,d)\right),
\label{equ: unified_formulation}
\end{equation} 
where $\psi$ and $\phi$  are representation functions that extract features from $q$ and $d$, respectively, $\eta$ is the interaction function that extracts features from the pair $(q,d)$, and $g$ is an evaluation function that computes the relevance score based on the feature representations. 
For neural IR models, in most cases, the representation functions $\psi$, $\phi$, $\eta$, and $g$ are encoded in the network structure so that all can be learned from training data \cite{guo2020deep}.

The ranking performance of an \ac{IR} model is usually evaluated by a metric $M$ that focuses on the top-$K$ ranking results, e.g., Recall@$K$, NDCG@$K$ and MRR@$K$. 
Given a triple $\left(q,D,Y\right)$ and an IR model $f$, the score of $M$ on query $q$ is calculated by:
\begin{equation}
    M\left(f ; \left(q,D,Y\right), K \right) = \sum_{(d, y_d) \in (D, Y)} y_d \cdot h\left(\pi_f\left(q, d\right)\right) \cdot \mathbb{I}\left\{\pi_f\left(q, d\right)\leq K\right\},
\label{equ: ranking_metric}
\end{equation} 
where $h$ is the mapping function related to ranking, dependent on the specific metric, and $\mathbb{I}\left\{\cdot\right\}$ is an indicator function which is equal to 1 when its condition is satisfied and 0 otherwise.

Further, given a test dataset $\mathcal{D}_\mathrm{test}$, the ranking performance $\mathcal{R}_M$ against metric $M$ is calculated by:
\begin{equation}
    \mathcal{R}_M\left(f ; \mathcal{D}_\mathrm{test}, K \right) = \frac{1}{\left|\mathcal{D}_\mathrm{test}\right|} \sum_{\left(q,D,Y\right) \in \mathcal{D}_\mathrm{test}} M\left(f ; \left(q,D,Y\right), K \right).
\label{equ: ranking_performance}
\end{equation} 
Typically, the ranking performance $\mathcal{R}_M$ of the IR model $f$ on the test dataset $\mathcal{D}_\mathrm{test}$ against metric $M$ refers to the average evaluation score of $M$.

\heading{IR models}
In order to balance efficiency and effectiveness, the IR task is usually addressed as a pipeline consisting of a first-stage retrieval stage and re-ranking stage \cite{chen2017efficient}:
\begin{enumerate}[label=(\arabic*)]
\item The first-stage retrieval identifies a small set of candidate documents from millions of documents.
Therefore, in the first-stage retrieval, $D_i$ refers to the entire corpus.
Considering efficiency, the neural IR model, represented by the \acf{DRM} \cite{guo2022semantic}, usually adopts a dual-encoder architecture.
In this architecture, the interaction function $\eta$ is often null.
\item The re-ranking stage generates the final ranked list for a query and a small set of candidate documents~\cite{guo2020deep}, referred to as $D_i$ in this stage. 
To this end, the \acf{NRM} with cross-encoder architecture is often modeled jointly by all the matching functions.
\end{enumerate}

\subsection{Definition of Robustness in IR}
Robustness refers to the ability to withstand disturbances or external factors that may cause a system to malfunction or provide inaccurate results \cite{huber1981robust}.
It is important for practical applications, especially in safety-critical scenarios, e.g., medical retrieval \cite{averbuch2004context}, financial retrieval \cite{hering2017annual}, and private information retrieval \cite{chor1998private}. 
If, for any reason, the \ac{IR} system behaves abnormally, the service provider can lose time, manpower, opportunities and even credibility.
With the development of deep learning, robustness has received much attention in the fields of \ac{CV} \cite{bhojanapalli2021understanding} and \ac{NLP} \cite{Wang2023TowardsAR}.
Concerns about model robustness in these fields are mainly focused on the test phase.
In this scenario, the model is trained on an unperturbed dataset but tested for its performance when exposed to adversarial examples or OOD data \cite{GoodfellowSS14, Quionero2009}.

In IR, the robustness of model in the test phase is also important due to the widespread availability of search engine optimization (SEO) \cite{gyongyi2005web} and the need for models to adapt to unseen data.
Hence, in this survey, we follow prior work and only discuss the robustness of a model in the test phase.

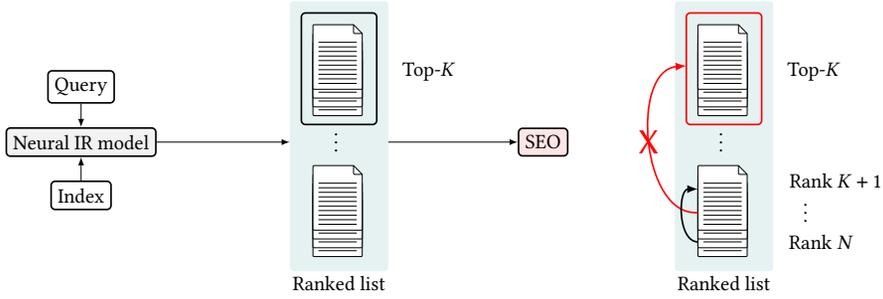
\begin{figure}[t]
    \centering
    \def\corner{0.15cm}
\def\cornerradius{0.02cm}
\def\lwidth{0.01cm}
\def\h{1.1cm}
\def\w{0.85cm}
\def\nline{10}
\def\iconmargin{0.1cm}
\def\topmargin{0.3cm}
\newcommand{\stackofpapers}{%
     \foreach[count=\i] \filename in {bar.c,bar.h,foo.c,foo.h}
     {
     \coordinate (nw) at ($(0.0cm*\i,0.15cm*\i)$);
     \coordinate (ne0) at ($(nw) + (\w, 0)$);
     \coordinate (ne1) at ($(ne0) - (\corner, 0)$);
     \coordinate (ne2) at ($(ne0) - (0, \corner)$);
     \coordinate (se) at ($(ne0) + (0, -\h)$); 
     \filldraw [-, line width = \lwidth, fill=white] (nw) -- (ne1) -- (ne2)
      [rounded corners=\cornerradius]--(se) -- (nw|-se) -- cycle;
     \draw [-, line width = \lwidth] (ne1) [rounded corners=\cornerradius]-- (ne1|-ne2) -- (ne2);
     \node [anchor=north west] at (nw) {\if0\scriptsize \tt \filename\fi};
     \foreach \k in {1,...,\nline}
     {
       \draw [-, line width = \lwidth, line cap=round] 
         ($(nw|-se) + (\iconmargin,\iconmargin) + (0,{(\k-1)/(\nline-1)*(\h - \iconmargin - \topmargin)})$)
           -- ++ ($(\w,0) - 2*(\iconmargin,0)$);
     }
     }
}
\resizebox{0.85\textwidth}{!}{%
\begin{tikzpicture}     
     \node[draw,rounded corners=2,thick,fill = gray!10] (A) at (-4,-2){Neural IR  model};
     \node[draw,rounded corners=2,thick, above = 0.4cm of A] (B) {Query};
     \node[draw,rounded corners=2,thick, below = 0.4cm of A] (C) {Index};
     \draw [-latex] (B) -- (A);
     \draw [-latex] (C) -- (A);
     \draw[draw=none,rounded corners=2,fill = teal!10] (-0.4,-4.2) rectangle (1.3,0.4) node at (0.45,-4.4) {Ranked list} node at (0.4,-1.9) {$\vdots$} ;
     \draw[rounded corners=2,thick] (-0.2,-1.7) rectangle (1.1,0.2) node at (2.0,-0.8) {Top-$K$} ;
     \begin{scope}[yshift=-0.6cm]	
       \stackofpapers
     \end{scope}
     \begin{scope}[yshift=-3.0cm]
       \stackofpapers
     \end{scope} 
     \draw[-latex]  (-2.70,-2) -- (-0.4,-2) ;     
     \node[draw,rounded corners=2,thick,fill = red!10] (E) at (3.97,-2){SEO};
     \draw[-latex]  (1.3,-2) -- (E.west) ;
     \draw[draw=none,rounded corners=2,fill = teal!10] (6.24,-4.2) rectangle (7.94,0.4) node at (7.09,-4.4) {Ranked list} node at (7.04,-1.9) {$\vdots$} ;     
     \draw[rounded corners=2,thick, color=red] (6.44,-1.7) rectangle (7.74,0.2) node at (8.64,-0.8) {\textcolor{black}{Top-$K$}} ;
     
     \begin{scope}[xshift=6.64cm, yshift=-0.6cm]	
     \stackofpapers
     \end{scope}
     \begin{scope}[xshift=6.64cm, yshift=-3.0cm]
     \stackofpapers     
     \end{scope}        
   \draw[thick, -latex, red] (6.64,-3.2) to [bend left=90] (6.44,-0.7) node at (5.79,-2) {\sffamily\huge X} ;
   \draw[thick, -latex] (6.64,-3.7) to [bend left=90] (6.64,-2.8) node at (9.02,-2.65) {Rank $K+1$} node at (8.5,-3.1) {$\vdots$} node at (8.77,-3.7) {Rank $N$} ;     
\end{tikzpicture}
}
    \caption{The core of robust IR is to protect the stability of the Top-K results.}
    \label{fig:Top-K}
\end{figure}

In most deployed systems, when presented with a ranked list users focus most of their attention on the top-$K$ search results, as evidenced by a significant drop in traffic and click-through rates further down the list \cite{niu2012top,wu2022certified} and by the prevalence of ranking metrics like MRR \cite{craswell2009mean} and NDCG \cite{jarvelin2002cumulated}, which primarily evaluate the effectiveness of these top-ranking results. 
The relationship between Top-$K$ result stability and robust \ac{IR} is shown in Figure \ref{fig:Top-K}.
Taking SEO as an example, it aims to get its document displayed in the Top-$K$-ranked results. 
And a robust neural IR model protects its Top-$K$ results from being affected.
Consequently, ensuring the integrity and robustness of the top-$K$-ranked results is crucial for deploying IR models into practical web search applications.
Based on this, we present a formal definition of \emph{top-$K$ robustness in IR} by incorporating the original test dataset $\mathcal{D}_\mathrm{test}$ from the initial dataset and the unseen test dataset $\mathcal{D}^*_\mathrm{test}$. 

\begin{definition}[Top-$K$ robustness in information retrieval] 
\label{def: robustness}
\rm
Let $\delta \geq 0$ denote an acceptable error threshold.
Let $f_{\mathcal{D}_\mathrm{train}}$ be  an IR model trained on training dataset $\mathcal{D}_\mathrm{train}$, with a corresponding test dataset $\mathcal{D}_\mathrm{test}$, and an unseen test dataset $\mathcal{D}^*_\mathrm{test}$, for the top-$K$ ranking results. 
If 
\begin{equation} 
    \left\lvert \mathcal{R}_M\left(f_{\mathcal{D}_\mathrm{train}} ; \mathcal{D}_\mathrm{test}, K \right)  - \mathcal{R}_M\left(f_{\mathcal{D}_\mathrm{train}} ; \mathcal{D}^*_\mathrm{test}, K \right) \right\rvert \leq \delta,
\end{equation}
we consider the model $f_{\mathcal{D}_\mathrm{train}}$ to be \emph{$\delta$-robust for metric $M$}. 
\end{definition}

\noindent%
In Section~\ref{Taxonomy}, below, $\mathcal{D}^*_\mathrm{test}$ refers to different test datasets depending on the context: $\mathcal{D}'_\mathrm{test}$ for test datasets with adversarial examples in adversarial robustness, $\tilde{\mathcal{D}}_\mathrm{test}$ for test dataset from new domains in OOD robustness.

\begin{figure}[t]
    \centering
    \includegraphics[width=\linewidth]{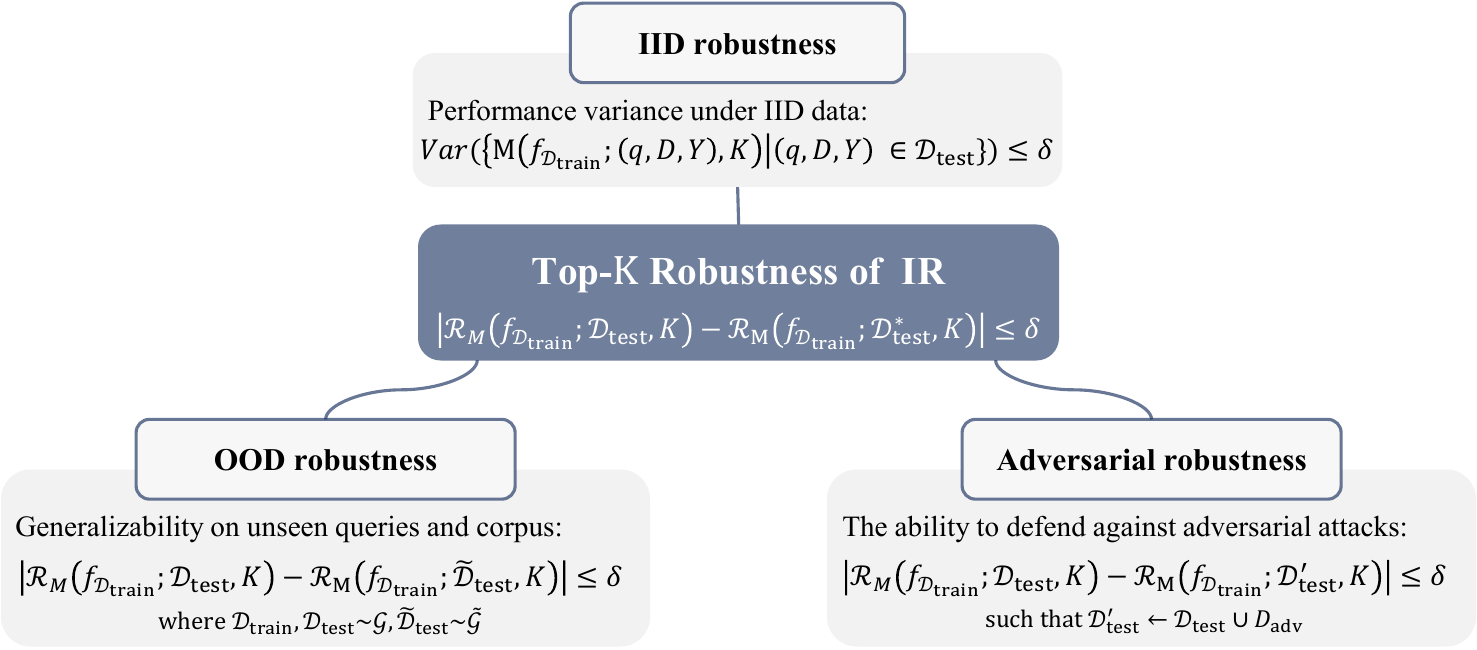}
    \caption{A taxonomy of robustness in IR. In this survey, we pay special attention to adversarial robustness and OOD robustness.}
    \label{fig:robust IR framework}
\end{figure}

\subsection{Taxonomy of Robustness in IR}
\label{Taxonomy}
In IR, robustness threats come in different flavors, including IID robustness \cite{wu2022neural}, OOD robustness~\cite{thakur2beir}, and adversarial robustness \cite{wu2022neural}. 
A high-level taxonomy of robustness in \ac{IR} is shown in Figure~\ref{fig:robust IR framework}.
 
\subsubsection{IID robustness}  
Typically, the performance of IR models is first represented by their overall performance on IID data.
Recently, it has been recognized that performance stability across IID queries may be compromised when we try to improve the average retrieval effectiveness across all queries \cite{zhang2013bias}.
Therefore, a robust neural IR model should not only have good retrieval performance on the overall testing queries, but also ensure that the performance on individual queries is not too bad.
We give a formal definition of IID robustness in IR based on Definition~\ref{def: robustness}.

\begin{definition}[IID robustness of information retrieval] 
\label{def: performance Variance}
\rm
Let the following be given: an IR model $f_{\mathcal{D}_\mathrm{train}}$ trained on training dataset $\mathcal{D}_\mathrm{train}$ with a corresponding IID test dataset $\mathcal{D}_\mathrm{test}$, and an acceptable error threshold $\delta$, for the top-$K$ ranking result. If
\begin{equation} 
     \operatorname{Var}\left( \{ M\left(f_{\mathcal{D}_\mathrm{train}}; \left(q, D, Y\right), K\right) \mid \left(q, D, Y\right) \in \mathcal{D}_{\text{test}} \} \right) \leq \delta, 
\end{equation}
where $\mathrm{Var}(\cdot)$ is the variance of the ranking performance of the \ac{IR} model $f_{\mathcal{D}_\mathrm{train}}$ on $\mathcal{D}_\mathrm{test}$, then the model $f$ is considered to be \emph{$\delta$-robust in terms of IID data for metric $M$}.
\end{definition}

\subsubsection{OOD robustness}
IR models need to cope with a constant stream of unseen data \cite{wu2022neural}.
The key behind addressing this challenge is how to adapt the model to new data out of the familiar distribution.
There are a variety of \ac{OOD} scenarios in \ac{IR}, so the \ac{OOD} robustness of the model is of broad interest.
First, the query entered by the user may be unknown and of varying quality \cite{penha2022evaluating,liu2023robustness}.
Then, search engine application scenario migration and incremental new documents will likely bring in \ac{OOD} data \cite{cai2023l2r,chen2023continual,thakur2beir}.
Manual labeling of unseen data as well as retraining IR models incurs significant resource overhead \cite{nguyen2016ms,thakur2beir}.
Therefore, it is crucial how to efficiently train \ac{IR} models to achieve effective performance on unseen data.

\ac{OOD} robustness measures the performance of an IR model on unseen queries and documents from distributions that differ from the training dataset.
By introducing the test dataset $\tilde{\mathcal{D}}_\mathrm{test}$ in a new domain, we give a formal definition of OOD robustness in IR based on Definition~\ref{def: robustness}.

\begin{definition}[OOD robustness of information retrieval] 
\label{def: ood robustness}
\rm
Let the following be given: an IR model $f_{\mathcal{D}_\mathrm{train}}$, an original dataset with training and test data, $\mathcal{D}_\mathrm{train}$ and $\mathcal{D}_\mathrm{test}$, drawn from the original distribution $\mathcal{G}$, along with a new test dataset $\tilde{\mathcal{D}}_\mathrm{test}$ drawn from the new distribution $\tilde{\mathcal{G}}$, and an acceptable error threshold $\delta$, for the top-$K$ ranking result. If
\begin{equation} 
    \left\lvert \mathcal{R}_M\left(f_{\mathcal{D}_\mathrm{train}} ; \mathcal{D}_\mathrm{test}, K \right)  - \mathcal{R}_M\left(f_{\mathcal{D}_\mathrm{train}} ; \tilde{\mathcal{D}}_\mathrm{test}, K \right) \right\rvert \leq \delta \text{ where } \mathcal{D}_\mathrm{train}, \mathcal{D}_\mathrm{test} \sim \mathcal{G}, \tilde{\mathcal{D}}_\mathrm{test} \sim \tilde{\mathcal{G}},
\end{equation}
the model $f$ is considered to be \emph{$\delta$-robust against OOD data for metric $M$}.
\end{definition}

\subsubsection{Adversarial robustness}
In a competitive scenario, content providers may aim to promote their products or documents in rankings for specific queries \cite{kurland2022competitive}.
This has provided a market for \ac{SEO} and has led to the development of attack techniques against search engines.
Traditional attacks against search engines are generally called \emph{term spamming}.
They usually resort to stacking keywords to achieve a ranking boost.
In recent years, with the development of deep learning, a number of neural approaches have emerged that attack through more imperceptible perturbations.
As search engines evolve, defense methods against term spamming are maturing as well. 
Adversarial robustness focuses on the stability of the IR model performance when imperceptible malicious perturbations are added to documents.
By introducing a test dataset $\mathcal{D}^{\prime}_\mathrm{test}$ with adversarial examples, we give a formal definition of adversarial robustness in IR based on Definition~\ref{def: robustness}.

\begin{definition}[Adversarial robustness in information retrieval] 
\label{def: adversarial robustness}
\rm
Let the following be given: an IR model $f_{\mathcal{D}_\mathrm{train}}$ trained on training dataset $\mathcal{D}_\mathrm{train}$ with a corresponding testing dataset $\mathcal{D}_\mathrm{test}$, a new document set $D_\mathrm{adv}$ containing  adversarial examples, and an acceptable error threshold $\delta$, for the top-$K$ ranking result. If
\begin{equation} 
    \left\lvert \mathcal{R}_M\left(f_{\mathcal{D}_\mathrm{train}} ; \mathcal{D}_\mathrm{test}, K \right)  - \mathcal{R}_M\left(f_{\mathcal{D}_\mathrm{train}} ; \mathcal{D}^{\prime}_\mathrm{test}, K \right) \right\rvert \leq \delta  \text{ such that } \mathcal{D}^{\prime}_\mathrm{test} \leftarrow \mathcal{D}_\mathrm{test} \cup D_\mathrm{adv} ,
\end{equation}
where $\mathcal{D}_\mathrm{test} \cup D_\mathrm{adv}$ denotes injecting the set of all generated adversarial examples $D_\mathrm{adv}$ into the original test dataset, then model $f$ is considered to be \emph{$\delta$-robust against adversarial examples for metric $M$}.
\end{definition}

\noindent%
Relatively little work has been done on IID robustness. 
Research on this type of robustness mainly includes \cite{wu2022neural,zhang2013bias,zhang2014bias,zhang2022bias,wang2009mean}.
In this paper, we pay special attention to the two other notions of robustness depicted in Figure~\ref{fig:robust IR framework}, i.e., adversarial robustness and OOD robustness. 
Existing work on adversarial robustness usually proceeds along two lines: adversarial attacks and adversarial defenses. 
We will discuss these lines in detail in Section~\ref{Sec: Adversarial Robustness}.
Depending on the types of \ac{OOD} generalizability in IR models, the existing work can be categorized into the OOD generalizability on unseen documents and the OOD generalizability on unseen queries.
We will discuss these directions in detail in Section~\ref{Sec: Out-of-distribution Robustness}.

\section{Adversarial Robustness} \label{Sec: Adversarial Robustness}
Deep neural networks have been found to be vulnerable to adversarial examples that can produce misdirection with human-imperceptible perturbations \cite{IanGoodfellow2014ExplainingAH,JavidEbrahimi2017HotFlipWA}. 
In IR, deep learning-based models are also likely to inherit these adversarial vulnerabilities~\cite{SzegedyZSBEGF13}, which raises concerns about the robustness of neural IR systems. 
Building on the definitions of adversarial robustness in IR provided in Section \ref{Taxonomy}, we survey work on adversarial robustness, focusing specifically on adversarial attacks and adversarial defenses.

\subsection{Overview}
In IR, search engine optimization (SEO) has been around since the dawn of the world wide web  \cite{gyongyi2005web}.
This includes white-hat SEO \cite{goren2020ranking}, which modifies documents in good faith and within the rules and expectations of search engines to optimize the quality of web pages.
In contrast, black-hat SEO \cite{castillo2011adversarial}, maliciously exploiting loopholes of search engines, is used to get a site ranking higher in search results. 


\paragraph{Traditional web spamming}
SEO involves the creation of web pages with no real informational value, designed to manipulate search engine algorithms and distort search results. This deceptive practice lures web users to sites they would not typically visit, undermining the trust relationship between users and search engines and potentially damaging the search engine's reputation \cite{baeza1999modern}.

\paragraph{Web spamming attack.}
When assessing textual relevance, search engines scrutinize the locations on a web page where query terms appear \cite{baeza1999modern}. 
These locations are referred to as fields. Common text fields for a page include the document body, the title, the meta tags in the HTML header, and the page's URL. 
Additionally, the anchor texts associated with URLs that point to the page are considered part of the page (anchor text field), as they often provide a succinct description of the page's content. 
The terms in these text fields play a crucial role in determining the page's relevance to a specific query, with different weights typically assigned to different fields. 
Term spamming refers to the manipulation of these text fields to make spam pages appear relevant for certain queries \cite{castillo2011adversarial}.

\paragraph{Web spamming detection.} 
Web spamming posed a serious threat to search engines in the early days.
However, they are also easy to spot and suspect due to their simple implementation \cite{gyongyi2005web}.
Web spamming detection involves identifying and mitigating the practice of term spamming, where keywords are excessively used or manipulated on a web page to artificially inflate its relevance to specific search queries \cite{robertson1994some}. 
Various techniques have been developed to detect term spamming.

There is a utility-based term spamicity detection method known as OSD (online spam detection) \cite{zhou2008spamicity, zhou2009osd} that has been used to identify adversarial examples. 
OSD mainly relies on TF-IDF features and has been validated by the Microsoft adCenter \cite{liu2022order}.
It introduces the notion of spamicity to measure how likely a page is spam. 
Spamicity is a more flexible and user-controllable measure than the traditional supervised classification methods.
Using spamicity, online link spam, and term spam can be efficiently detected. 
However, this detection method can only target changes that add query terms and is prone to misclassification.

\paragraph{Why study adversarial attacks in IR\@?}
With the development of deep learning, neural networks are beginning to be widely used in IR models and have achieved excellent performance.
Although traditional web spamming can also have a significant attack effect on neural IR models, this method does not really pose a threat due to its ease of detection.
However, in the context of black-hat SEO, neural IR models are at risk of being attacked due to the inherent vulnerability inherited from neural networks.
Therefore, adversarial attacks are beginning to be studied to expose vulnerability flaws in neural IR models in advance.

\begin{figure}[t]
    \centering
    \resizebox{0.7\textwidth}{!}{%
   \begin{tikzpicture}
     \def\corner{0.15cm}
     \def\cornerradius{0.02cm}
     \def\lwidth{0.01in}
     \def\h{1.1cm}
     \def\w{0.85cm}
     \def\nline{10}
     \def\iconmargin{0.1cm}
     \def\topmargin{0.3cm}
     \node[draw,rounded corners=2,thick,fill = red!10] (A) at (-4.8,0){Attack};
     \node[draw,rounded corners=2,thick,fill = green!10] (D) at (5,0){Defense};
     \draw[black,rounded corners=2,thick,fill = teal!10] (-0.8,-0.82) rectangle (0.92,1) 
     	node at (0.1,1.3) {Top-$K$ results} ;
     \begin{scope}[yshift=.8cm]
     \foreach[count=\i] \filename in {bar.c,bar.h,foo.c,foo.h}
     {
     \coordinate (nw) at ($(-0.15cm*\i,-0.075cm*\i)$);
     \coordinate (ne0) at ($(nw) + (\w, 0)$);
     \coordinate (ne1) at ($(ne0) - (\corner, 0)$);
     \coordinate (ne2) at ($(ne0) - (0, \corner)$);
     \coordinate (se) at ($(ne0) + (0, -\h)$); 
     \filldraw [-, line width = \lwidth, fill=white] (nw) -- (ne1) -- (ne2)
      [rounded corners=\cornerradius]--(se) -- (nw|-se) -- cycle;
     \draw [-, line width = \lwidth] (ne1) [rounded corners=\cornerradius]-- (ne1|-ne2) -- (ne2);
     \node [anchor=north west] at (nw) {\if0\scriptsize \tt \filename\fi};
     \foreach \k in {1,...,\nline}
     {
       \draw [-, line width = \lwidth, line cap=round] 
         ($(nw|-se) + (\iconmargin,\iconmargin) + (0,{(\k-1)/(\nline-1)*(\h - \iconmargin - \topmargin)})$)
           -- ++ ($(\w,0) - 2*(\iconmargin,0)$);
     }
     }
     \end{scope}
     \draw[-Stealth]  (-4.16,0) -- (-0.8,0) node [pos=0.5,above] {change} node [pos=0.5,below] {identify flaws};
     \draw[-Stealth]  (4.29,0) -- (0.93,0) node [pos=0.5,above] {stabilize\vphantom{g}} node [pos=0.5,below] {strengthen models};
   \end{tikzpicture}
}
    \caption{Purpose and relationship between adversarial attacks and defenses.}
    \label{fig:attack-defense}
\end{figure}
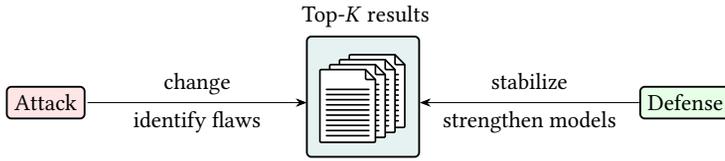

\paragraph{Why study adversarial defense in IR\@?}
To mitigate adversarial attacks, there is a growing body of work that is focusing on adversarial defenses.
Adversarial defense focuses on the early hardening of model vulnerabilities discovered by adversarial attacks.
Its goal is to obtain robust neural retrieval models to build reliable IR systems.

The relationship between adversarial attacks and defenses is shown in Figure \ref{fig:attack-defense}.
Recently, adversarial robustness has begun to gain widespread attention.
Below, we describe these efforts from several perspectives: benchmark datasets, adversarial attacks, and adversarial defenses.

\begin{table}[t]
  \caption{Benchmark datasets for studying adversarial robustness. \#Documents denotes the number of documents in the corpus; \#Q$_{\mathrm{train}}$ denotes the number of queries available for training; \#Q$_{\mathrm{dev}}$ denotes the number of queries available for development; \#Q$_{\mathrm{eval}}$ denotes the number of queries available for evaluation.}
  \label{tab:attack datasets}
  \small
  \begin{tabular}{l rrrr l}
    \toprule
   \textbf{Dataset} &\textbf{\#Documents} & \textbf{\#Q$_{\mathrm{train}}$} & \textbf{\#Q$_{\mathrm{dev}}$} & \textbf{\#Q$_{\mathrm{eval}}$} & \textbf{References} \\
    \midrule
    MS MARCO document \cite{nguyen2016ms} & 3.2M & 370K & 5,193  & 5,793 & \cite{wu2023prada,liu2022order,liu2023black,liu2024multi,wu2022certified} \\
    MS MARCO passage \cite{nguyen2016ms} & 8.8M & 500K & 6,980  & 6,837 & \cite{wang2022bert,zhong2023poisoning,wu2023prada,liu2023black,liu2024perturbation,wu2022certified,chen2023towards} \\
    Clueweb09-B \cite{clarke2009overview} & 50M & 150 & --  & -- &  \cite{wang2022bert,liu2024multi} \\
    Natural Questions \cite{kwiatkowski2019natural} & 21M & 60K & 8.8K  & 3.6K & \cite{liu2022order,long2024backdoor,zhong2023poisoning} \\
    TriviaQA \cite{joshi2017triviaqa} & 21M & 60K & 8.8K & 11.3K & \cite{long2024backdoor} \\
    \midrule
    TREC DL19 \cite{Craswell_Mitra_Yilmaz_Campos_Voorhees_2020} & -- & -- & 43  & -- & \cite{liu2022order,wang2022bert,parry2024analyzing} \\
    TREC DL20 \cite{craswell2021overview} & -- & -- & 54 & -- & \cite{parry2024analyzing} \\
    TREC MB14 \cite{lin2013overview} & -- & -- & 50  & -- & \cite{liu2022order} \\
    \midrule
    ASRC \cite{raifer2017information} & 1,279 & -- & 31  & -- & \cite{wu2022neural,goren2018ranking} \\
    Q-MS MARCO \cite{liu2023topic} & -- & -- & 4,000  & -- & \cite{liu2023topic} \\
    Q-Clueweb09 \cite{liu2023topic} & -- & -- & 292  & -- & \cite{liu2023topic} \\
    DARA \cite{chen2023defense} & 164k & 50k &  3,490 & 3,489 & \cite{chen2023defense} \\
  \bottomrule
\end{tabular}
\end{table}

\subsection{Benchmark Datasets}
In this section, we present the datasets commonly used for studying adversarial robustness. 
Prior work on attacks and defenses against robustness has focused on experiments on previusly published IR datasets, as shown in Table \ref{tab:attack datasets}. 
All datasets can be found in the BestIR benchmark; see Appendix~\ref{appendix:B} for an overview of the resources collected in BestIR.

\paragraph{Basic datasets}
Some datasets in IR are adapted for reuse by attack and defense methods as basic datasets.
These include MS MARCO document/passage \cite{nguyen2016ms} and Clueweb-09B \cite{clarke2009overview}. 
Some work \cite{wu2023prada,liu2023black,liu2024multi} performed experiments against attacks and defenses directly on these datasets.
For example, prior work usually used the training set of the basic dataset to train an NRM as the target model \cite{wu2023prada,liu2024multi}; the queries in the development set are used as target queries, and a portion of the documents in the ranked list of each query is sampled as target documents.
Attacks on these target documents measure the performance of adversarial attack methods.
We will discuss the specific evaluation methods in Section \ref{Sec 3.3 Adversarial Attack}.

\paragraph{Expansion of datasets}
Some collaborative benchmarks, such as TREC DL19 \cite{Craswell_Mitra_Yilmaz_Campos_Voorhees_2020} and TREC DL20 \cite{craswell2021overview}, have provided additional query collections for evaluation against the base dataset. 
Similarly, these query collections can be used to evaluate the effectiveness of attack methods.
Queries in these datasets are often attacked as additional sets of targeted queries.
For example, existing work usually trains an NRM on the MS MARCO passage dataset and uses the 43 queries in TREC DL19 as the target queries to perform attacks \cite{chen2023towards,liu2022order}. 

\paragraph{Off-the-shelf datasets}
Some research has adapted the above datasets to construct new datasets that can be used directly to perform attacks or evaluate defenses.
For example, ASRC \cite{raifer2017information} is based on documents in Clueweb, which are manually modified to generate new adversarial samples for evaluating the model's adversarial defense abilities.
Existing work also used it to study the effects of manual manipulation of documents on IR systems \cite{vasilisky2023content,goren2018ranking}.
To perform a topic-oriented attack, \citet{liu2023topic} constructed query groups on the same topic based on ORCAS \cite{craswell2020orcas} and the TREC 2012 Web Track \cite{clarke2009overview} as a complement to MS MARCO document and Clueweb-09b, respectively.
DARA \cite{chen2023defense} is a dataset for detecting adversarial ranking attacks and includes two types of detection task for adversarial documents.

\subsection{Adversarial Attacks} \label{Sec 3.3 Adversarial Attack}
As neural networks have become increasingly prevalent in IR systems, they have also become a target for adversarial attacks. 
Studying adversarial attacks can help understand the vulnerability of neural IR models before deploying them in real-world applications, and it can also be used as a surrogate evaluation and support the development of appropriate countermeasures.

\begin{algorithm}[t]
 \caption{Adversarial sample generation for neural IR models}
 \label{attack alg}
 \begin{algorithmic}[1] 
 \REQUIRE \quad \\
 A target query $q$, a target document $d$, a query collection $\mathcal{Q}$, a corpus $\mathcal{C}$, a neural IR model \\
 $f$, and a ranking loss function $\tilde{\mathcal{L}}$
 \ENSURE An adversarial document $d^{adv}$ \\
 \IF{$f$ is a black-box model}
 \STATE \textbf{Procedure} Surrogate model imitation
 \STATE Query the model $f$ with $\mathcal{Q}$, and collect ranked lists.
 \STATE Train the surrogate model $\tilde{f}$ with $\mathcal{Q}$ and the collected ranked lists.
 \STATE $f = \tilde{f}$
 \ENDIF\\[1mm]
 \STATE \textbf{Procedure} Adversarial attack
 \STATE Initialize the adversarial example $d^{adv}$ as a copy of the target document $d$.
 \FOR{$t\leftarrow 1$ to $\eta$} 
 \STATE Query the model $f$ with the target query $q$, and collect the ranked list $D$.
 \STATE Calculate the gradient of $\tilde{\mathcal{L}}$ with respect to the target query $q$ and target document $d$:
 \STATE $\operatorname{gradient} \leftarrow \nabla_{d}\tilde{\mathcal{L}}\left(f,q,d^{adv},D\right)$
 \STATE Generate the higher dimensional adversarial perturbation $\rho$:
 \STATE $\rho \leftarrow \operatorname{normalize}(\operatorname{gradient})$
 \STATE Mapping high-dimensional perturbations to text space:
 \STATE $p \leftarrow  \rho$
 \STATE Add textual perturbations:
 \STATE $d^{adv} \leftarrow d \oplus p$
 \ENDFOR
 \RETURN $d^{adv}$
 \end{algorithmic}
\end{algorithm}

\subsubsection{What are the differences between IR attacks and CV/NLP attacks?}
Adversarial attacks have undergone significant development in the fields of CV and NLP \cite{JavidEbrahimi2017HotFlipWA,IanGoodfellow2014ExplainingAH}, where attacks are typically directed at image retrieval and text classification tasks, respectively. 
However, the landscape of adversarial attacks differs in IR:
\begin{enumerate*}[label=(\roman*)]
\item Compared with image retrieval attacks, the IR attacks need to maintain semantic consistency of the perturbed document with the original document by considering the textual semantic similarity, rather than pixel-level perturbations within a fixed range in continuous space; and
\item Inspired by black-hat SEO, the goal of IR attacks is to inject imperceptible perturbations within a document to improve its ranking for one or multiple specific queries within the entire candidate set or corpus, not inducing classification errors by the model. 
\end{enumerate*}

Without loss of generality, given a query $q$ and a target document $d$, the goal of generating imperceptible perturbations $p$ to attack against a neural IR model $f$ under top-$K$ ranked results can usually be formalized as:
\begin{equation}
\max_{p} \; \left(K - \pi_f\left(q, d \oplus p\right) + \lambda \cdot \operatorname{Sim}\left(d, d \oplus p\right)   \right) ,
\end{equation}
where $\pi_f\left(q, d \oplus p\right)$ denotes the ranking position of the perturbed document $d\oplus p$ in the ranked list generated by $f$ with respect to query $q$.  
The $\operatorname{Sim}\left( \cdot \right)$ function measures the similarity between the adversarial example and the original document, both textually as well as semantically.
$\lambda$ is a regularization parameter used to balance two goals: keeping the adversarial samples as close as possible to the original document, while allowing adversarial samples to be ranked as high as possible.
Ideally, the adversarial sample $d \oplus p$ should preserve the original semantics of $d$,  and be imperceptible to human judges yet misleading to the neural IR models.

\subsubsection{What is the attack setting?}
Depending on whether the attacker has access to the knowledge of the parameters of the target model, the attack setup can be categorized into two main types~\cite{wu2023prada,liu2022order}: 
\begin{enumerate}[label=(\arabic*)]
\item \textbf{White-box attacks}: Here, the attacker can fully access the model parameters and use the model gradient to directly generate perturbations.
\item \textbf{Black-box attacks}: Here, the model parameters cannot be obtained; the attacker usually adopts a transfer-based black-box attack paradigm \cite{cheng2019improving}:
They construct a surrogate white-box model by continuously querying the model and getting the output. 
Specifically, a surrogate model is trained to simulate the performance of the target model, and then the surrogate model is attacked to the transferability of the adversarial samples to indirectly attack the target model. 
In IR, attackers can query the target model to obtain a ranked list with rich information.
Therefore, the surrogate model often has access to sufficient training data, making this attack effective \cite{wu2023prada,liu2022order}. 
\end{enumerate}
There are various attack methods and target models in IR.
We present pseudo-code to illustrate a fundamental IR attack in Algorithm \ref{attack alg}.

\begin{figure}[t]
    \centering
    \input{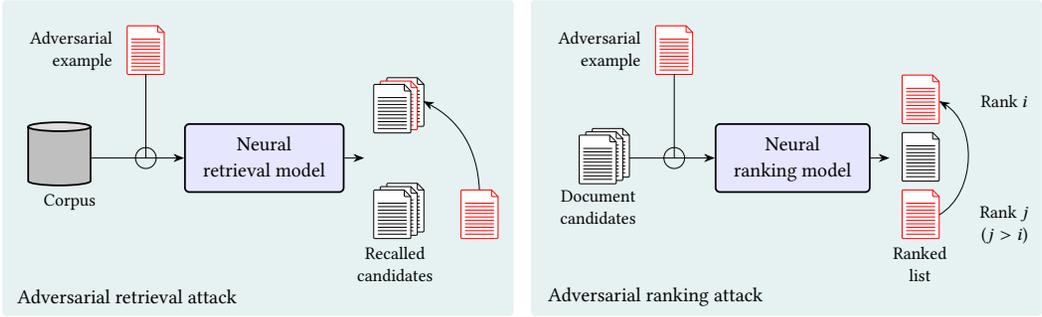}
    \caption{Adversarial retrieval attacks vs. adversarial ranking attacks.}
    \label{fig:retrieval vs. ranking}
\end{figure}

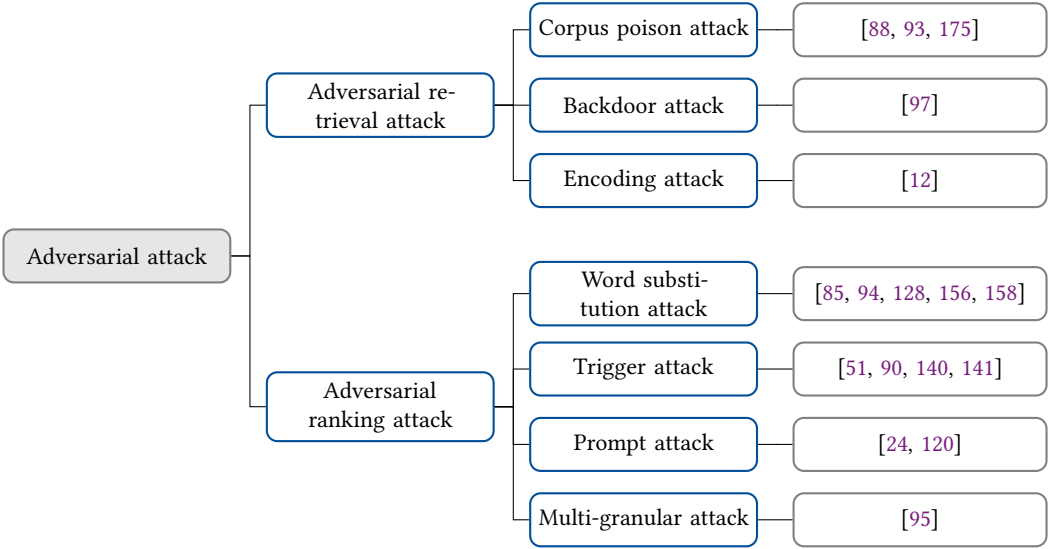
\begin{figure}[h] 
  \centering 
    \resizebox{1.0\textwidth}{!}{%
    \begin{tikzpicture}[
      grow = right,
      level 1/.style={level distance=2cm, sibling distance=4cm}, 
      level 2/.style={level distance=2cm, sibling distance=1cm}, 
      level 3/.style={level distance=2cm, sibling distance=1cm}, 
      edge from parent path = {
        (\tikzparentnode.east) -| +(0.25,0) |- (\tikzchildnode.west)
      },
      bag/.style={
        text width = 8em,
        text centered,
        anchor = west,
        fill = white,
        draw = inkblue,
        thick,
        rounded corners,
        font=\small,
        minimum height = 2em},
      cite/.style={
        text width = 9em,
        text centered,
        anchor = west,
        fill = white,
        draw = gray,
        thick,
        rounded corners,
        font=\small,
        minimum height = 2em},
      ]

      \node[bag, fill = gray!20, draw = gray] {Adversarial attack} 
          child  {
            node[bag] {Adversarial ranking attack}
              child {
                node [bag] {Multi-granular attack}
                  child {
                    node [cite] {\cite{liu2024multi}}
                  }
              }
              child {
                node [bag] {Prompt attack}
                  child {
                    node [cite] {\cite{chen2023towards,parry2024analyzing}}
                  }
              }
              child {
                node [bag] {Trigger attack}
                  child {
                    node [cite] {\cite{song2020adversarial,goren2020ranking,liu2022order,song2022trattack}}
                  }
              }
              child {
                node [bag] {Word substitution attack}
                  child {
                    node [cite] {\cite{raval2020one,liang2018deep,wang2022bert,wu2023prada,liu2023topic}}
                  }
              }
          }
          child  {
            node[bag] {Adversarial retrieval attack}
              child {
                node [bag] {Encoding attack}
                  child {
                    node [cite] {\cite{boucher2023boosting}}
                  }
              }
              child {
                node [bag] {Backdoor attack}
                  child {
                    node [cite] {\cite{long2024backdoor}}
                  }
              }
              child {
                node [bag] {Corpus poison attack}
                  child {
                    node [cite] {\cite{liu2023black,zhong2023poisoning,lin2023mawseo}}
                  }
              }
          }
      ;
    \end{tikzpicture}%
    }
  \caption{Classification of adversarial attacks against neural IR models.} 
  \label{fig:attack-tree} 
\end{figure}

According to the type of target model, attack efforts against IR models can be broadly categorized into two types, which are visualized in Figure~\ref{fig:retrieval vs. ranking}:
\begin{enumerate}[label=(\arabic*)]
    \item \textbf{Adversarial retrieval attacks} target the first-stage retrieval, mainly against dense retrieval models; and
    \item \textbf{Adversarial ranking attacks} target the re-ranking stage, mainly against neural ranking models. 
\end{enumerate}
The methodology for conducting adversarial attacks in IR can differ based on the chosen attack strategy and the target model.
A categorization of adversarial attacks in IR is shown in Figure \ref{fig:attack-tree}.

Next, we introduce evaluation, adversarial retrieval attacks, and adversarial ranking attacks of adversarial attacks in IR.

\subsubsection{Evaluation}
The evaluation of adversarial attacks includes both attack performance and naturalness performance.

\paragraph{Attack performance} 
Attack performance mainly refers to the degree of ranking improvement after a target document has been attacked. 
In general, the following automatic metrics for attack performance are widely adopted:  

\begin{itemize}
\item \textbf{Attack success rate} (ASR/SR) \cite{wu2023prada,liu2022order,liu2023topic}, which evaluates the percentage of target documents successfully boosted under the corresponding target query;

\item \textbf{Average boosted ranks} (Boost/Avg.boost) \cite{liu2023black,liu2022order,liu2023topic}, which evaluates the average improved rankings for each target document under the corresponding target query;

\item \textbf{Boosted top-$K$ rate} (T$K$R) \cite{liu2022order,liu2023topic}, which evaluates the percentage of target documents that are boosted into top-$K$ w.r.t.\ the corresponding target query; and

\item \textbf{Normalized ranking shifts rate} (NRS) \cite{wang2022bert,liu2023black}, which evaluates the relative ranking improvement of adversarial examples that are successfully recalled into the initial set with $K$ candidates:
\begin{equation}
\operatorname{NRS} = ({\Pi_d - \Pi_{d^{adv}}})/{\Pi_d} \times 100 \%,
\end{equation}
where $\Pi_d$ and $\Pi_{d^{adv}}$ are the rankings of $d$ and $d^{adv}$ respectively, produced by the target IR model. 
\end{itemize}

The effectiveness of an adversary is better with a higher value for all these metrics. 

\paragraph{Naturalness performance} 
Naturalness performance refers primarily to the degree to which a target document is imperceptible to humans after it has been attacked.
In general, the following automatic metrics for naturalness performance and human evaluation are widely adopted:  
\begin{itemize}
\item \textbf{Automatic spamicity detection}, which can detect whether target pages are spam or not; the utility-based term spamicity method OSD \cite{zhou2009osd} is usually used to detect the adversarial examples; 

\item \textbf{Automatic grammar checkers}, i.e., Grammarly\footnote{\url{https://app.grammarly.com/}} and Chegg Writing,\footnote{\url{https://writing.chegg.com/}} which calculate the average number of errors in the adversarial examples; 

\item \textbf{Language model perplexity} (PPL), which measures the fluency using the average perplexity calculated using a pre-trained GPT-2 model~\cite{radford2019language}; and

\item \textbf{Human evaluation}, which measures the quality of the attacked documents w.r.t. aspects of imperceptibility, fluency, and semantic similarity \cite{liu2023topic,liu2022order,wu2022neural}. 
\end{itemize}

\subsubsection{Adversarial retrieval attack}
In this subsection, we introduce the task definition of adversarial retrieval attacks and methods to achieve such attacks. 

\paragraph{Task definition} 
The adversarial retrieval attack task is designed for attacks against DRMs.
The objective of adversarial retrieval attacks is centered around the manipulation of a document that initially fails to be recalled. 
By integrating adversarial perturbations, the aim is to ensure that this document is subsequently retrieved by the first-stage neural retrieval model, thereby securing its presence within the candidate set. 
This approach not only challenges the robustness and reliability of neural retrieval systems but also raises significant questions regarding the integrity of information authenticity \cite{liu2023black,zhong2023poisoning}. 
The goal of adversarial retrieval attacks under top-$K$ ranked results can be formalized as:
\begin{equation}
\max_{p} \; \left(K - \operatorname{Recall}_f\left(q, d \oplus p\right) + \lambda  \cdot\operatorname{Sim}\left(d, d \oplus p\right)  \right) ,
\end{equation}
where $\operatorname{Recall}_f\left(q, d \oplus p\right)$ denotes the recalled position of the perturbed document $d \oplus p$ produced by the first stage retrieval model $f$ with respect to query $q$ given the entire corpus.
A smaller value of $\operatorname{Recall}$ denotes a higher ranking.

\paragraph{Method}
Current retrieval attack methods mainly include corpus poison attacks, backdoor attacks, and encoding attacks.

\begin{itemize}
\item \textbf{Corpus poison attack.} 
Corpus poisoning attacks construct adversarial samples against a specific query and inject them into the corpus in the inference phase so that they are recalled.
MCARA \cite{liu2023black} addresses this issue and attempts to mine the vulnerability of DRMs.
It introduces the adversarial retrieval attack (AREA) task, which intends to deceive a DRM into retrieving a target document that is outside the initial set of candidate documents.
\citet{zhong2023poisoning} adopt the HotFlip method \cite{JavidEbrahimi2017HotFlipWA} from NLP to iteratively add perturbations in the discrete token space to maximize its similarity to a set of queries.
In this way, they maximize the contamination of the corpus by a limited number of documents.
MAWSEO \cite{lin2023mawseo} employs adversarial revisions to achieve real-world cybercriminal objectives, including rank boosting, vandalism detection evasion, topic relevancy, semantic consistency, user awareness (but not alarming) of promotional content, etc. 

\item \textbf{Backdoor attack.} 
Backdoor attacks inject a small proportion of ungrammatical documents into the corpus. 
When user queries contain grammatical errors, the model will recall the learned triggering pattern and assign high relevance scores to those documents.
\citet{long2024backdoor} introduces a novel scenario where the attackers aim to covertly disseminate targeted misinformation, such as hate speech or advertisements, through a retrieval system. 
To achieve this, they propose a backdoor attack triggered by grammatical errors and ensure that attack models can function normally for standard queries but are manipulated to return passages specified by the attacker when users unintentionally make grammatical mistakes in their queries.

\item \textbf{Encoding attack.} 
By imperceptibly perturbing documents using uncommon encoded representations, encoding attacks control results across search engines for specific search queries.
\citet{boucher2023boosting} make words look the same as they originally do by adding an offset encoding to them, while the search engines are deceived.
The experiment on a mirror of Simple Wikipedia shows that the proposed method can successfully deceive search engines in realistic scenarios.
\end{itemize}

\subsubsection{Adversarial ranking attack}
In this subsection, we introduce the task definition of adversarial ranking attacks and methods to achieve adversarial ranking attacks. 

\paragraph{Task definition} 
The adversarial ranking attack task is designed to attack against NRMs.
Adversarial ranking attacks typically involve introducing adversarial perturbations to a document already present in the candidate set, to manipulate its ranking position either elevating or diminishing it as determined by a neural ranking model.
The goal of adversarial ranking attacks under top-$K$ ranked results can usually be formalized as:
\begin{equation}
\max_{p} \;\left(K - \operatorname{Recall}_f\left(q, d \oplus p\right) + \lambda  \cdot\operatorname{Sim}\left(d, d \oplus p\right) \right) ,
\end{equation}
where $\operatorname{Rank}_f\left(q, d \oplus p\right)$ denotes the position of the perturbed document $d \oplus p$ in the final ranked list generated by the NRM $f$ with respect to query $q$.
A smaller value of $\operatorname{Rank}$ denotes a higher ranking.


\heading{Method}
Adversarial ranking attacks against NRMs include word substitution attacks, trigger attacks, and prompt attacks. 

\begin{itemize}
\item \textbf{Word substitution attack.} 
Word substitution attacks typically boost a document's ranking by replacing a small number of words in the document with synonyms.
A common method of \emph{white-box} word substitution attack is the gradient-based attack, where the attacker uses the gradient of the loss function with respect to the input data to create adversarial examples. 
These examples are designed to cause the model to make incorrect relevance predictions or rankings.
\citet{raval2020one} present a systematic approach of using adversarial examples to measure the robustness of popular ranking models. 
They follow an approach that is similar to one used in text classification tasks \cite{liang2018deep} and perturb a limited number of tokens (with a minimum of one) in documents, replacing them with semantically similar tokens such that the rank of the document changes.
Brittle-BERT \cite{wang2022bert} adds/replaces a small number of tokens to a highly relevant or non-relevant document to cause a large rank demotion or promotion.
The authors find a small set of recurring adversarial words that, when added to documents, result in successful rank demotion/promotion of any relevant/non-relevant document, respectively. \

As for \emph{black-box} word substitution attacks, in the field of machine learning, \citet{SzegedyZSBEGF13} find that adversarial examples have the property of cross-model transferability, i.e., the adversarial example generated by a surrogate model can also fool a target model.
Black-box attacks in information retrieval usually adopt this transfer-based paradigm due to the excellent performance of imitation of the target model.
\citet{wu2023prada} propose the first black-box adversarial attack task against NRMs, the word substitution ranking attack (WSRA) task.
The WSRA task aims to fool NRMs into promoting a target document in rankings by replacing important words in its text with synonyms in a semantic-preserving way.
Based on this task, authors propose a novel pseudo relevance-based adversarial ranking attack method (PRADA), which outperformed web spamming methods by 3.9\% in ASR.
The WSRA task focuses only on attacks on single query-document pairs and does not take into account the dynamic nature of search engines.
Based on this, \citet{liu2023topic} introduce the topic-oriented adversarial ranking attack (TARA) task, which aims to find an imperceptible perturbation that can promote a target document in ranking for a group of queries with the same topic. 

\item \textbf{Trigger attack.} 
Trigger attacks boost document rankings by injecting a generated trigger sentence into a specific location in the document (e.g., the beginning).
\citet{song2020adversarial} propose using semantically irrelevant sentences (semantic collisions) as perturbations.
They develop gradient-based approaches for generating collisions given white-box access to an NRM.
\citet{goren2020ranking} propose a document manipulation strategy to improve document quality for the purpose of improving document ranking.
\citet{liu2022order} propose a trigger attack method, PAT, empowered by a pairwise objective function, to generate adversarial triggers, which causes premeditated disorderliness with very few tokens. 
TRAttack \cite{song2022trattack} uses rewriting existing sentences in the text to improve document ranking with learning ability from the multi-armed bandit mechanism.

\item \textbf{Prompt attack.} 
Prompt attacks use prompts to guide a language model to generate perturbations based on existing documents to improve document ranking.
\citet{chen2023towards} propose a framework called imperceptible document manipulation (IDEM) to produce adversarial documents that are less noticeable to both algorithms and humans. 
IDEM finds the optimal connect sentence to insert into the document through a language model.
\citet{parry2024analyzing} analyze the injection of query-independent prompts, such as ``true'' into documents and find that the prompt perturbation method is valid for several sequence-to-sequence relevance models like monoT5 \cite{nogueira2020document}.

\item \textbf{Multi-granular attack.} 
Multi-granular attacks focus on generating high-quality adversarial examples by incorporating multi-granular perturbations, i.e., word level, phase level, and sentence level.
\citet{liu2024multi} propose RL-MARA, a reinforcement learning framework, to navigate an appropriate sequential multi-granular ranking attack path.
By incorporating word-level, phrase-level, and sentence-level perturbations to generate imperceptible adversarial examples, RL-MARA is able to increase the flexibility of creating adversarial examples, thereby improving the potential threat of the attack. 
\end{itemize}

\subsection{Adversarial Defenses}
With the advent of SEO, many defenses have been created to counter malicious attacks.
In the field of adversarial defenses, much work has been devoted to training robust neural IR models or identifying malicious adversarial examples in advance.

\subsubsection{IR defense task}
The primary objective of defenses in IR is to maintain, or even enhance, the performance of IR models when the test dataset includes adversarial examples. 
This involves the implementation of strategies during the training or inference phases: the goal is to ensure that the model's ability to accurately retrieve relevant documents remains uncompromised, even in the presence of manipulative adversarial perturbations.

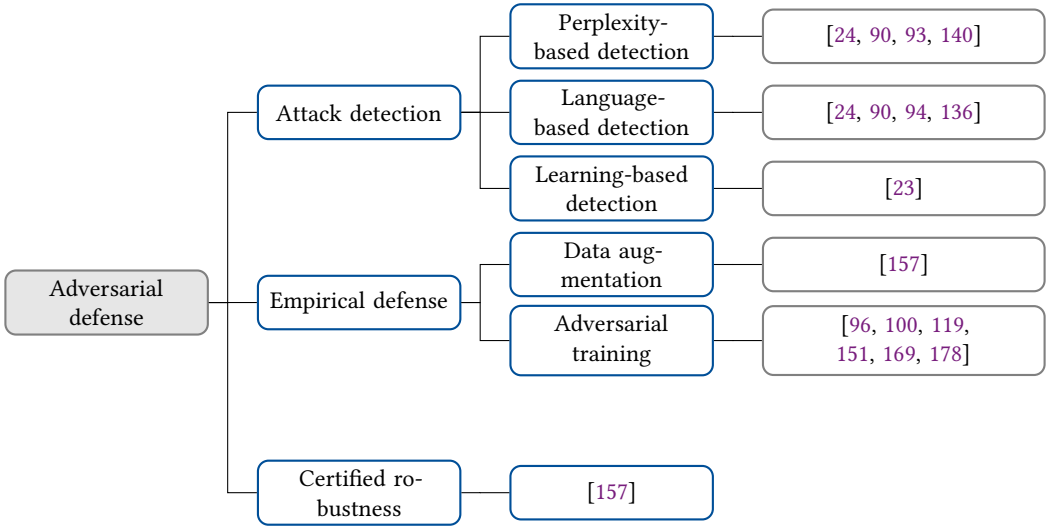
\begin{figure}[t] 
  \centering 
    \begin{tikzpicture}[
      grow = right,
      level 1/.style={level distance=2cm, sibling distance=2.5cm}, 
      level 2/.style={level distance=2cm, sibling distance=1cm}, 
      level 3/.style={level distance=2cm, sibling distance=1cm}, 
      edge from parent path = {
        (\tikzparentnode.east) -| +(0.25,0) |- (\tikzchildnode.west)
      },
      bag/.style={
        text width = 7em,
        text centered,
        anchor = west,
        fill = white,
        draw = inkblue,
        thick,
        rounded corners,
        font=\small,
        minimum height = 2em},
      cite/.style={
        text width = 10em,
        text centered,
        anchor = west,
        fill = white,
        draw = gray,
        thick,
        rounded corners,
        font=\small,
        minimum height = 2em},
      ]

      \node[bag, fill = gray!20, draw = gray] {Adversarial defense} 
          child  {
            node[bag] {Certified robustness}
              child {
                node [bag] {\cite{wu2022certified}}
              }
          }
          child  {
            node[bag] {Empirical defense}
              child {
                node [bag] {Adversarial training}
                  child {
                    node [cite] {\cite{wang2017irgan, zhang2021adversarial, zhou2023towards,park2019adversarial,lupart2023study,liu2024perturbation}}
                  }
              }
              child {
                node [bag] {Data augmentation}
                  child {
                    node [cite] {\cite{wu2022certified}}
                  }
              }
          }
          child  {
            node[bag] {Attack detection}
              child {
                node [bag] {Learning-based detection}
                  child {
                    node [cite] {\cite{chen2023defense}}
                  }
              }
              child {
                node [bag] {Language-based detection}
                  child {
                    node [cite] {\cite{shen2023textdefense, chen2023towards,liu2022order,liu2023topic}}
                  }
              }
              child {
                node [bag] {Perplexity-based detection}
                  child {
                    node [cite] {\cite{song2020adversarial,chen2023towards,liu2022order,liu2023black}}
                  }
              }
          }
      ;
    \end{tikzpicture}%
  \caption{Classification of adversarial defenses in neural IR.} 
  \label{fig:defense-tree} 
\end{figure}

Without loss of generality, given a test set $\mathcal{D}_\mathrm{test}$ and an adversarial document set $D_\mathrm{adv}$, the goal of adversarial defense against an neural IR model $f$ under top-$K$ ranked results can usually be formalized as:
\begin{equation} 
    \label{robustness definition}
    \max \mathcal{R}_M\left(f_{\mathcal{D}_\mathrm{train}} ; \mathcal{D}^{\prime}_\mathrm{test}, K \right) \text{ such that } \mathcal{D}^{\prime}_\mathrm{test} \leftarrow \mathcal{D}_\mathrm{test} \cup D_\mathrm{adv} .
\end{equation}
The adversarial defense task is in the training or testing phase.
In the testing phase, it is usually in the form of attack detection.
The training phase is usually in the form of both empirical defense and certified robustness.
We present pseudo code to illustrate a fundamental IR defense, as shown in Algorithm \ref{defense alg}.
A detailed categorization of adversarial defense in \ac{IR} is shown in Figure \ref{fig:defense-tree}.

\begin{algorithm}[t]
\caption{Adversarial Defense in Neural IR Models}
\label{defense alg}
\begin{algorithmic}[1] 
\REQUIRE \quad \\
A query collection $\mathcal{Q}$, a corpus $\mathcal{C}$, a neural IR model $f$, potentially adversarial documents 
\ENSURE Safe and robust document rankings

\STATE \textbf{Procedure} Attack Detection
\STATE Train a detector model $g$ using benign and adversarial examples from $\mathcal{Q}$ and $\mathcal{C}$
\FOR{each document $d \in \mathcal{C}$}
    \STATE Compute the probability $g(d)$ of $d$ being adversarial
    \IF{$g(d) > \text{threshold}$}
        \STATE Flag document $d$ as adversarial
    \ENDIF
\ENDFOR\\[1mm]

\STATE \textbf{Procedure} Empirical Defense
\STATE Fine-tune $f$ on adversarial examples using augmented training set $\mathcal{Q}^{\prime}$, $\mathcal{C}^{\prime}$
\FOR{each training iteration}
    \STATE Apply random transformations to documents in $\mathcal{C}^{\prime}$ to simulate adversarial perturbations
    \STATE Update $f$ to minimize loss on transformed documents
\ENDFOR\\[1mm]

\STATE \textbf{Procedure} Certified Robustness
\STATE Incorporate certified defense methods into $f$ (e.g., randomized smoothing)
\FOR{each query $q \in \mathcal{Q}$}
    \STATE Compute a robustness certificate for the ranking produced by $f(q)$
    \IF{certificate fails}
        \STATE Adjust $f$ to enhance robustness for $q$
    \ENDIF
\ENDFOR

\RETURN Updated and defended neural IR model $f$
\end{algorithmic}
\end{algorithm}

Next, we introduce the evaluation, attack detection, empirical defense, and certified robustness of adversarial defense in IR.

\subsubsection{Evaluation}
Adversarial defense assessment includes metrics for the training phase and metrics for the inference phase.
Specifically, defenses for the training phase mainly comprise of empirical defenses and certified robustness; and defenses for the inference phase mainly concern the detection of adversarial samples.

\paragraph{Metrics used in the training phase} 
The metrics in the training phase are mainly for evaluating the ability of empirical defenses and certified robustness methods to maintain the original ranked list in the presence of adversarial samples:

\begin{itemize}
    \item \textbf{CleanMRR@$k$} evaluates Mean Reciprocal Rank (MRR) \cite{craswell2009mean} performance on a clean dataset;
    \item \textbf{RobustMRR@$k$} \cite{liu2024perturbation} evaluates the MRR performance on the attacked test dataset;
    \item \textbf{Attack success rate} (ASR) \cite{liu2024perturbation} evaluates the percentage of the after-attack documents that are ranked higher than the original documents; and 
    \item \textbf{Location square deviation} (LSD) \cite{liu2024perturbation} evaluates the consistency between the original and perturbed ranked list for a query, by calculating the average deviation between the document positions in the two lists. 
\end{itemize}

\paragraph{Metrics used in the inference phase} 
The metrics in the inference phase are mainly used for evaluating the ability of attack detection methods to accurately recognize adversarial samples:

\begin{itemize}
    \item \textbf{Point-wise detection accuracy} \cite{chen2023defense} evaluates the correctness of the detection of whether a single document has been perturbed or not;

    \item \textbf{\#DD} \cite{chen2023defense} denotes the average number of discarded documents ranked before the relevant document; and

    \item \textbf{\#DR} \cite{chen2023defense} denotes the average number of discarded relevant documents.
\end{itemize}

\subsubsection{Attack detection}
While progress in empirical defenses and certified robustness aids in training NRMs to be more robust in their defense against potential attacks, the detection of adversarial documents has also been explored.

\paragraph*{Perplexity-based detection}
Perplexity-based detection mainly uses the difference in the distribution of perplexity between the adversarial samples and the original document under the language model for recognition.
Adversarial perturbations applied to original documents can significantly impact the semantic fluency of their content \cite{liu2022order, liu2023black, chen2023towards}. 
\citet{song2020adversarial} have developed a perplexity-based detection to counter ranking attacks. 
Detection involves using a pre-trained language model (PLM) to assess the perplexity of documents, where higher perplexity values indicate less fluent text. 
Consequently, any document surpassing a certain perplexity threshold is filtered out from consideration.

\paragraph*{Language-based detection}  
Language-based detection primarily uses unnatural language representations to identify adversarial samples.
Adversarially generated or modified documents often exhibit grammatical inconsistencies or lack context coherence \cite{shen2023textdefense, chen2023towards}, since documents that have been maliciously perturbed are usually grammatically incorrect and incoherent.
Some work uses grammar checkers, i.e., Grammarly and Chegg Writing, to detect adversarial examples \cite{liu2022order,liu2023topic}.
The number of grammatical errors in the target document according to the grammar checkers and the quality scoring are used as indicators of linguistic acceptability.
Any document deemed to have poor linguistic acceptability is subsequently discarded.

\paragraph*{Learning-based detection} 
Learning-based detection uses neural networks to model the characteristics of adversarial samples, and empirically learns to distinguish adversarial samples from clean samples.
As mentioned earlier, spamming, perplexity-based, and language-based detectors lack knowledge of the adversarial documents, potentially leading to sub-optimal performance. 
Consequently, \citet{chen2023defense} introduce two kinds of detection task, namely point-wise and list-wise detection, to standardize evaluation processes of the efficacy of adversarial ranking detection methods.
They fine-tune the BERT and RoBERTa models using the original and adversarial document pairs present in the training set of the generated dataset.
Experimental results demonstrate that a supervised classifier can effectively mitigate known attacks, the detection accuracy can be up to 99.5\%, but it performs poorly against unseen attacks.

\subsubsection{Empirical defense}
Empirical defenses attempt to make models empirically robust to known adversarial attacks; this has been extensively explored in image retrieval \cite{madry2018towards} and text classification \cite{wang2021adversarial}.
The aim of an empirical defense is to find adversarial examples during training and use them to augment the training set. 

\paragraph{Data augmentation}
Data augmentation often employs randomized or heuristic methods to transform the training samples, thereby expanding the training data.
In the training phase, models that have seen augmented data will have some defense against adversarial samples.
There are many data augmentation methods in NLP that achieve good defense performance \cite{ribeiro2018semantically, jia2017adversarial}.

In IR, \citet{wu2022certified} have been the first to apply data augmentation to the defense of NRMs.
They find that data augmentation can reduce the attack success rate to some extent. However, it performs worse than customized defenses for IR.
Hence, simply augmenting the training documents (as in NLP) is not a robust defense against attacks in IR.

\paragraph{Adversarial training}
Adversarial training is one of the most effective defenses against specific seen attacks. 
By integrating pre-constructed adversarial samples into training data, adversarial training has demonstrated a strong defense in both CV and NLP \cite{madry2018towards,zhu2019freelb}.

In IR, there is a body of work that implements adversarial training by means of adversarial optimization \cite{wang2017irgan, zhang2021adversarial, zhou2023towards}, such as GAN \cite{goodfellow2020generative}. 
In this context, the goal is often to simply improve the effectiveness of the IR model.
Meanwhile, there is work by \citet{lupart2023study} and \citet{park2019adversarial} who attempt to use adversarial training for improving robustness.
By incorporating adversarial examples into training data, adversarial training has become the de facto defense approach to adversarial attacks against NRMs. 
However, this defense mechanism is subject to a trade-off between effectiveness and adversarial robustness. 

To tackle this issue, \citet{liu2024perturbation} theoretically define the perturbation invariance of a ranking model and design a perturbation-invariant adversarial training (PIAT) method for ranking models to achieve a better effectiveness-robustness trade-off. 
Experimental results on several ranking models demonstrate the superiority of PIAT compared to earlier adversarial defenses. 

\subsubsection{Certified robustness}
Since empirical defenses are only effective for certain attacks rather than all attacks, competition emerges between adversarial attacks and defense methods.
To solve the attack-defense dilemma, researchers resort to certified defenses to make models provably robust to certain kinds of adversarial perturbations. 
In NLP, \citet{jia2019certified} have been the first to propose to certify the robustness of adversarial word substitutions by using interval bound propagation \citep[IBP,][]{dvijotham2018training}. 
 
In the field of IR, \citet{wu2022certified} propose a rigorous and provable certified defense method for NRMs.
They define certified Top-$K$ robustness for ranking models since users mainly care about the top-ranked results in real-world scenarios. 
Then, they propose CertDR to achieve certified top-$K$ robustness, based on the idea of randomized smoothing. 
Their experiments demonstrate that CertDR can significantly outperform state-of-the-art empirical defense methods for ranking models.

\section{OOD Robustness} \label{Sec: Out-of-distribution Robustness}
In addition to adversarial vulnerability, deep neural networks lack generalizability to OOD data. 
When faced with data that differs from the distribution of the training data, neural networks struggle to maintain performance. 
In IR, this problem has begun to be exposed and attract attention as neural IR models are now widely being used \cite{wu2022neural,thakur2beir}.
Prior work focuses on the OOD generalizability of DRMs, since OOD data has a direct impact on the retrieval stage.
We refer back to Section~\ref{Taxonomy} for a definition of OOD robustness in IR.
In this section, we present specific work on OOD generalizability on unseen documents and  OOD generalizability on unseen queries, respectively.

\begin{figure}[t]
    \centering
    \resizebox{\textwidth}{!}{%
	\begin{tikzpicture}     
		\draw [dashed] (0,-3) -- (0,-7.75);
		\draw[draw=none,rounded corners=2,fill = teal!10] (-7.5,-7.75) rectangle (-3.5,-4.2) ;
		\node[draw=none,rounded corners=2,thick, fill=none] at (-5.55,-7.45) {Adapt to OOD corpus};
		\draw[draw=none,rounded corners=2,fill = teal!10] (3.55,-7.75) rectangle (7.55,-4.2) ;
		\node[draw=none,rounded corners=2,thick, fill=none] at (5.55,-7.45) {Adapt to OOD query set};
	
		\node[draw,rounded corners=2,thick,fill = blue!10,yshift=-5cm] (A) at (0,0){Neural IR model};
          
		\node[cylinder, draw, shape aspect=.6, cylinder uses custom fill, cylinder end fill=red!10, 
		minimum height=1cm, cylinder body fill=red!20, opacity=1, scale=1, 
		rotate=90, minimum width=1cm, thick, left = 2cm of A.west, yshift=1.03cm,xshift=.582cm] (B) {};     

		\node[cylinder, draw, shape aspect=.6, cylinder uses custom fill, cylinder end fill=gray!10, 
      		minimum height=1cm, cylinder body fill=gray!20, opacity=1, scale=1, 
      		rotate=90, minimum width=1cm, thick, above = 1cm of B, xshift=.4cm, yshift=-1.03cm] (C) {};

		\node[draw=none,rounded corners=2,thick, left = 0.75 cm of C, fill=none, 
		yshift=.5cm] (D) {Training corpus};

		\node[cylinder, draw, shape aspect=.6, cylinder uses custom fill, cylinder end fill=green!10, 
		minimum height=1cm, cylinder body fill=green!20, opacity=1, scale=1, 
		rotate=90, minimum width=1cm, thick, below = 1cm of B, yshift=1.03cm,xshift=-.6cm] (E) {};     

		\node[cylinder, draw, shape aspect=.6, cylinder uses custom fill, cylinder end fill=blue!10, 
		minimum height=1cm, cylinder body fill=blue!20, opacity=1, scale=1, 
		rotate=90, minimum width=1cm, thick, left = 1cm of B, yshift=1.13cm, xshift=1.02cm] (F) {};     

		\node[cylinder, draw, shape aspect=.6, cylinder uses custom fill, cylinder end fill=purple!10, 
		minimum height=1cm, cylinder body fill=purple!20, opacity=1, scale=1, 
		rotate=90, minimum width=1cm, thick, above = 1cm of F, yshift=-1.03cm, xshift=-2.5875cm] (G) {};     
		
		\node [rectangle, below of=B, xshift=-10.5mm, yshift=2.5mm] (split1) {};
     		\node[coordinate] (auxLeft) at (-0.5,-5.25) {};
		
		\draw [-Stealth] (C) -- (A.west);
		\draw [-Stealth,brown] (A) -- (B);
		\draw [-Stealth,brown,rounded corners=0.05cm] (auxLeft) --($(auxLeft)+(0,-1.34)$)-- (E);
     
     		\path [draw,brown,rounded corners=0.05cm] (auxLeft) |- (split1) ;
		\path [draw,brown,rounded corners=0.05cm, -Stealth] (split1) -- ($(split1)+(0.2,-0.0)$) -| (F) ;
		\path [draw,brown,rounded corners=0.05cm, -Stealth] (split1) -- ($(split1)+(0.2,-0.0)$) -| (G) ;

	\newcommand{\cube}[1]{%
    \coordinate (CenterPoint) at (0,0);
    \def\width{0.84cm};
    \def\height{0.84cm};
    \def\textborder{0.1cm};
    \def\xslant{0.2cm};
    \def\yslant{0.15cm};
    \def\rounding{0.2pt};
    \node[thick, draw,
          minimum height  = \height,
          minimum width   = \width,
          text width      = {\width-2*\textborder},
          align           = center,
          fill            = #1!20,
          rounded corners = \rounding]
          at (CenterPoint) {}; 
    \draw [rounded corners = \rounding, thick, fill=#1!10] %
        ($(CenterPoint) + (-\width/2. - 2*\rounding, \height/2.)$) -- %
        ($(CenterPoint) + (-\width/2. + \xslant - 2*\rounding, \height/2. + \yslant)$) -- %
        ($(CenterPoint) + (\width/2. + \xslant + 2*\rounding, \height/2. + \yslant)$) -- %
        ($(CenterPoint) + (\width/2. + 2*\rounding, \height/2.)$) -- %
        cycle;
    \draw [rounded corners = \rounding, thick, fill=#1!30] %
        ($(CenterPoint) + (\width/2. + \xslant + 2*\rounding, \height/2. + \yslant)$) -- %
        ($(CenterPoint) + (\width/2. + 2*\rounding, \height/2.)$) -- %
        ($(CenterPoint) + (\width/2. + 2*\rounding, -\height/2.)$) -- %
        ($(CenterPoint) + (\width/2. + \xslant + 2*\rounding, -\height/2. + \yslant)$) -- %
        cycle;
        }
        \begin{scope}[xshift=4.47cm,yshift=-3.61cm]
        \cube{gray}
        \end{scope}
        \begin{scope}[xshift=4.47cm,yshift=-5.01cm]
        \cube{red}
        \end{scope}
        \begin{scope}[xshift=4.47cm,yshift=-6.61cm]
        \cube{green}
        \end{scope}
        \begin{scope}[xshift=6.67cm,yshift=-5.01cm]
        \cube{blue}
        \end{scope}
        \begin{scope}[xshift=6.67cm,yshift=-6.61cm]
        \cube{purple}
        \end{scope}
        
       	\node[draw=none,rounded corners=2,thick, fill=none] at (6.7,-3.5) {Training queries};

	\node [coordinate] (split2) at (6.4,-5.748) {};
	\node[coordinate] (auxRight) at (0.5,-5.25) {};

	\node[coordinate] (aux1) at (4.05,-3.8) {};
	\draw [-Stealth] (aux1) -- (A.east);
	\node[coordinate] (aux2) at (4.05,-5.01) {};
	\draw [-Stealth,brown] (A.east) -- (aux2);
	\node[coordinate] (aux3) at (4.05,-6.61) {};
	\draw [-Stealth,brown,rounded corners=0.05cm] (auxRight) --($(auxRight)+(0,-1.34)$)-- (aux3);
		
     	\path [draw,brown,rounded corners=0.05cm] (auxRight) |- (split2) ;
	\node[coordinate] (aux4) at (6.67,-5.44) {};	
	\node[coordinate] (aux5) at (6.67,-6.03) {};	
	\path [draw,brown,rounded corners=0.05cm, -Stealth] (split2) -- ($(split2)+(0.2,-0.0)$) -| (aux4) ;
	\path [draw,brown,rounded corners=0.05cm, -Stealth] (split2) -- ($(split2)+(0.2,-0.0)$) -| (aux5) ;

   \end{tikzpicture}
}
    \caption{OOD generalizability on unseen documents vs. queries in neural IR.}
    \label{fig: OOD corpus queries}
\end{figure}
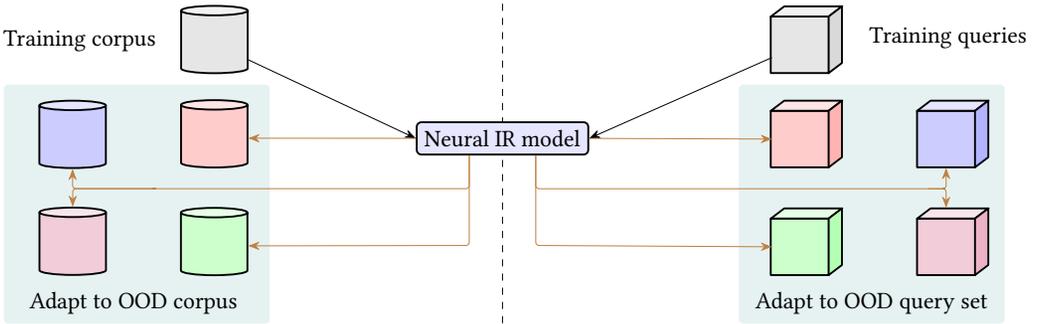

\subsection{Overview}
For a long time, research work on neural IR models has been carried out in a narrow independent and identically distributed (IID) setting.
In this setting, the model faces homogeneous data during training and testing.
But IR systems are widely used in fields such as search engines \cite{ilan1998search}, digital libraries \cite{chowdhury2003introduction}, medical search \cite{luo2008medsearch}, legal search \cite{maxwell2008concept}, and at the same time, the scenarios that IR systems need to cope with are becoming more complex.

\paragraph*{OOD generalizability requirements in IR}
With deep neural network models being applied to IR, neural IR models have demonstrated excellent results on many tasks \cite{guo2020deep,guo2022semantic}.
But IR systems need to face more than just a single task or scenario \cite{thakur2beir}.
In order to deploy an IR system, it is often necessary to construct a training dataset for the neural IR model in it.
This process is usually time-consuming and expensive and hence many IR systems are expected to be able to cope with a wide variety of data not seen during training \cite{thakur2beir,liu2023robustness,penha2022evaluating}.
Therefore, OOD generalizability is a key requirement for contemporary IR systems, given the dynamic nature of user needs and evolving data landscapes. 
In IR, OOD generalizability is focused on unseen documents and unseen queries as illustrated in Figure~\ref{fig: OOD corpus queries}.

\paragraph{Why should IR models be able to generalize to unseen documents\@?}
In real-world scenarios, the data landscape is constantly evolving, with new documents and information being generated regularly. 
It is expensive to annotate each new corpus and retrain the IR models.
Therefore, a neural IR model that can generalize well to unseen corpora ensures its relevance and usefulness over time, without requiring constant retraining or fine-tuning. 
Moreover, in complex real-life scenarios, generalizability to a new corpus helps IR models against distributional shifts or domain-specific biases. 
This helps to ensure that IR models deliver reliable ranking results irrespective of diverse contexts, qualities, and domains.

\paragraph{Why should IR models be able to generalize to unseen queries\@?}
The set of possible queries that users may input is vast and constantly evolving \cite{graus-2018-birth,penha2022evaluating,zhuang2021dealing}. 
E.g., 15\% of daily Google searches are brand new.\footnote{\url{https://blog.google/products/search/our-latest-quality-improvements-search/}}
The nature of information needs is dynamic and diverse \cite{guo2011intent,lefortier-2014-online}. 
Users often express their information needs in varied ways, using different vocabulary, language styles, or even typos \cite{zhuang2023typos,chen2022towards}. 
This challenge becomes particularly pronounced in the context of ever-changing user interests and the introduction of new vocabularies. 
Therefore, IR models must possess the ability to handle queries that were not encountered during training. 
Without generalizability to unseen queries, IR models risk providing inadequate or irrelevant results, ultimately diminishing user satisfaction and trust in the system \cite{penha2022evaluating,campos2023noise}.
A robust neural IR model should be able to understand and accommodate these variations, effectively retrieving relevant information regardless of how the query is formulated.

Furthermore, in reality, there is a wide variety of query types that are often not fully or adequately accessible during IR model training \cite{zhuang2022characterbert,zhuang2021dealing}.
But, a robust IR system should perform consistently in response to a wide range of query types.

\subsection{Benchmark Datasets}
In this section, we present the commonly used datasets for studying OOD robustness in IR. All datasets can be found in the BestIR benchmark; details about BestIR can be found in Appendix~\ref{appendix:B}.

\subsubsection{Datasets for unseen documents}
The datasets for unseen documents mainly involve adaptation to new corpus datasets and updates to corpus datasets.
Datasets on unseen documents in IR are listed in Table \ref{tab:corpus ood datasets}.

\begin{table}[t]
  \caption{Benchmark datasets for unseen documents. \#Documents denotes the number of documents in corpus; \#Q$_{\mathrm{train}}$ denotes the number of queries available for training; \#Q$_{\mathrm{dev}}$ denotes the number of queries available for development; \#Q$_{\mathrm{eval}}$ denotes the number of queries available for evaluation; \#Corpora denotes the number of corpora.}
  \label{tab:corpus ood datasets}
  \small
  \begin{tabular}{p{2.5cm}l rrrr c}
    \toprule
  \textbf{Type} & \textbf{Dataset} & \textbf{ \#Corpora} & \multicolumn{4}{c}{\textbf{Publications}} \\
    \midrule
  \begin{tabular}[c]{@{}l@{}}Adaptation to a\\ new corpus\end{tabular} & BEIR \cite{thakur2beir} & 18 & 
  \multicolumn{4}{m{5.2cm}}{\cite{Oh2023Data,Chen2023Cross-domain,Anaya-Isaza2022Data,laitz2024inranker,anand2023data,bonifacio2022inpars,li2024domain,ge2023augmenting,ram2022learning, dai2022promptagator,liu2022challenges,reddy2021towards,yu2022coco,izacard2021unsupervised,neelakantan2022text,liang2020embedding, ma2021zero,chandradevan2024duqgen,thakur2beir,ma2021zero,sachan2023questions,cai2022hyper,wang2022gpl,zhan2022disentangled,xian2023learning,xu2023berm,kang2024improving,xin2022zero,yu2022coco,ni2022large,lu2022ernie,huebscher2022zero,formal2022distillation,kasela2024desire,chen2022out,lee2023back}}
  \\
    \midrule
  \textbf{Type} & \textbf{Dataset} & \textbf{ \#Documents} & \textbf{\#Q$_{\mathrm{train}}$} & \textbf{\#Q$_{\mathrm{dev}}$} & \textbf{\#Q$_{\mathrm{eval}}$} & \textbf{Publications} \\
    \midrule
    \multirow{4}{*}{\begin{tabular}{@{}l}Updates to a\\ corpus\end{tabular}} 
    & CDI-MS \cite{chen2023continual} & 3.2M & 370K & 5,193  & 5,793 & \cite{chen2023continual} \\
    & CDI-NQ \cite{chen2023continual} & 8.8M & 500K & 6,980  & 6,837 & \cite{chen2023continual} \\
    & LL-LoTTE \cite{cai2023l2r} & 5.5M & 16K & 8.5K   & 8.6K  & \cite{cai2023l2r} \\
    & LL-MultiCPR \cite{cai2023l2r} & 3.0M & 136K & 15K  & 15K & \cite{cai2023l2r} \\
  \bottomrule
\end{tabular}
\end{table}

\paragraph{Adaptation to a new corpus}
In order to model the migration of models between corpora of different domains, datasets that cater for adaptation to a new corpus typically aggregate multiple domain-specific IR datasets.
Of these, BEIR \cite{thakur2beir} is the best-known example; it includes nine retrieval tasks, such as fact-checking, news retrieval, question answering, entity retrieval.
It also has 18 datasets, across diverse tasks, diverse domains, task difficulties, and diverse annotation strategies.

\paragraph{Updates to a corpus}
Updates to a corpus focus on the performance of an IR model when updates to a corpus occur.
In order to follow updates to a corpus over time, the available datasets are mainly constructed by slicing or expanding the existing dataset.
For example, to mimic the arrival of new documents in MS MARCO \cite{nguyen2016ms}, CDI-MS \cite{chen2023continual}  first randomly sampled 60\% documents from the whole corpus as the base documents. 
Then, it randomly samples 10\% documents from the remaining corpus as the new document set, which is repeated 4 times.

\subsubsection{Datasets for unseen queries}
The datasets for unseen queries mainly involve query variation datasets and unseen query type datasets.
Existing datasets on unseen queries in IR, are shown in Table \ref{tab:query ood datasets}.

\begin{table}[t]
  \caption{Benchmark datasets for unseen queries. \#Q$_{\mathrm{eval}}$ denotes the number of queries available for evaluation.}
  \label{tab:query ood datasets}
  \small
  \begin{tabular}{llrc}
    \toprule
  \textbf{Type} & \textbf{Dataset} & \textbf{\#Q$_{\mathrm{eval}}$} & \textbf{Publications} \\
    \midrule
   \multirow{9}{*}{Query variation} & DL-Typo \cite{zhuang2022characterbert} & 60 & \cite{zhuang2022characterbert} \\
   & noisy-MS MARCO \cite{campos2023noise} & 5.6k & \cite{campos2023noise} \\
   & rewrite-MS MARCO \cite{campos2023noise} & 5.6k & \cite{campos2023noise} \\
   & noisy-NQ \cite{campos2023noise} & 2k & \cite{campos2023noise} \\
   & noisy-TQA \cite{campos2023noise} & 3k & \cite{campos2023noise} \\
   & noisy-ORCAS \cite{campos2023noise} & 20k & \cite{campos2023noise} \\
   & variations-ANTIQUE \cite{penha2022evaluating} & 2k & \cite{penha2022evaluating} \\
   & variations-TREC19 \cite{penha2022evaluating} & 430 & \cite{penha2022evaluating} \\  
   & \citet{zhuang2021dealing} & 41k & \cite{zhuang2021dealing}
   \\
    \midrule
   \multirow{2}{*}{Unseen query type} & MS MARCO \cite{nguyen2016ms} & 15k & \cite{wu2022neural} \\
   & L4 \cite{surdeanu2008learning} & 10k & \cite{cohen2018cross} \\
  \bottomrule
\end{tabular}
\end{table}

\paragraph*{Query variation datasets}
The importance of query variation datasets lies in their ability to simulate real-world search scenarios, where users often have unique ways of expressing their information needs.
Query variation datasets contain sets of queries that target the same information need but are expressed in alternative ways, reflecting the natural diversity in how different users might phrase their search queries.
Such datasets can include paraphrased queries, queries with typos, order-swapped queries, and queries without stop words.
For example, \citet{penha2022evaluating} construct query variation datasets by turning queries from TREC DL19 \cite{Craswell_Mitra_Yilmaz_Campos_Voorhees_2020} and ANTIQUE \cite{hashemi2020antique} into different variants using 4 categories in 10 ways.

\paragraph{Unseen query type datasets}
Unseen query type datasets have queries that are not represented in the training data, either by virtue of their topic or the nature of the information being sought.
For example, the MS MARCO dataset \cite{nguyen2016ms} only contains 5 types of queries, i.e., location, numeric, person, description, and entity.
The primary purpose of these datasets is to test the generalization ability of IR models to novel, real-world query scenarios that users may present.

\subsection{OOD Generalizability to Unseen Documents} \label{sec: OOD corpus}
\begin{figure}[t] 
  \centering 
  \resizebox{1.0\textwidth}{!}{%
    \begin{tikzpicture}[
      grow = right,
      level 1/.style={level distance=2cm, sibling distance=4cm}, 
      level 2/.style={level distance=2cm, sibling distance=1.2cm}, 
      level 3/.style={level distance=2cm, sibling distance=1.2cm}, 
      edge from parent path = {
        (\tikzparentnode.east) -| +(0.25,0) |- (\tikzchildnode.west)
      },
      bag/.style={
        text width = 8em,
        text centered,
        anchor = west,
        fill = white,
        draw = inkblue,
        thick,
        rounded corners,
        font=\small,
        minimum height = 2em},
      cite/.style={
        text width = 9em,
        text centered,
        anchor = west,
        fill = white,
        draw = gray,
        thick,
        rounded corners,
        font=\small,
        minimum height = 2em},
      ]

      \node[bag, fill = gray!20, draw = gray] {OOD generalizability on unseen documents} 
          child  {
            node[bag] {Updates to a corpus}
              child {
                node [bag] {Continual learning for dense retrieval}
                  child {
                    node [cite] {\cite{cai2023l2r}}
                  }
              }
              child {
                node [bag] {Continual learning for generative retrieval}
                  child {
                    node [cite] {\cite{yoon2023continually,kishore2023incdsi,chen2023continual,guo2024corpusbrain++,mehta2023dsi++}}
                  }
              }
          }
          child  {
            node[bag] {Adaptation to a new corpus}
              child {
                node [bag] {Scaling up the model capacity}
                  child {
                    node [cite] {\cite{ni2022large,lu2022ernie}}
                  }
              }
              child {
                node [bag] {Architectural modifications}
                  child {
                    node [cite] {\cite{huebscher2022zero,formal2022distillation,kasela2024desire,chen2022out,lee2023back}}
                  }
              }
              child {
                node [bag] {Domain modeling}
                  child {
                    node [cite] {\cite{zhan2022disentangled,xian2023learning,xu2023berm,kang2024improving,xin2022zero,yu2022coco}}
                  }
              }
              child {
                node [bag] {Data augmentation}
                  child {
                    node [cite, font=\scriptsize] {\cite{Oh2023Data,Chen2023Cross-domain,Anaya-Isaza2022Data,laitz2024inranker,anand2023data,bonifacio2022inpars,li2024domain,ge2023augmenting,ram2022learning, dai2022promptagator,liu2022challenges,reddy2021towards,yu2022coco,izacard2021unsupervised,neelakantan2022text,liang2020embedding, ma2021zero,chandradevan2024duqgen,thakur2beir,ma2021zero,sachan2023questions,cai2022hyper,wang2022gpl}}
                  }
              }
          }
      ;
    \end{tikzpicture}%
    }
  \caption{Classification of OOD generalizability on unseen documents.} 
  \label{fig:corpus-tree} 
\end{figure}
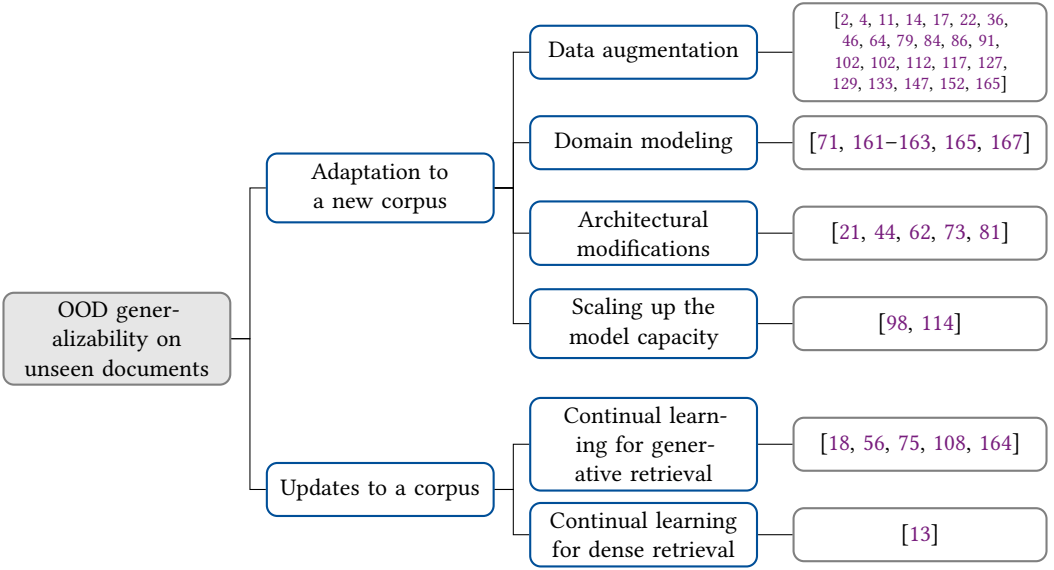

As argued above, IR systems need to adapt to different environments and variations in the corpus. 
However, retraining the neural IR models in each new environment is costly. 
Previous work has only analyzed the generalizability of IR models across different domains \cite{wu2022neural,thakur2beir,ren2023thorough}.
In this work, we summarize work on adaptation to a new corpus and updates to a corpus.
Figure \ref{fig:corpus-tree} illustrates how we organize the discussion of different methodologies.

\subsubsection{Definition}
Generalizability to unseen documents implies the ability of an IR model to maintain retrieval performance when encountering a new and unfamiliar corpus.
In IR, improving the model's OOD generalizability under unseen documents is mainly reflected in enhancing the retrieval performance of the IR model under various new corpus. 
Without loss of generality, given a test set $\tilde{\mathcal{D}}_\mathrm{test}$ with new corpus $\tilde{\mathcal{C}}$, they draw the new distribution $\tilde{\mathcal{G}}$, the goal of improving the OOD generalizability of a neural IR model $f$ on unseen documents under top-$K$ ranked results can usually be formalized as:
\begin{equation} 
    \label{OOD corpus}
    \max \mathcal{R}_M\left(f_{\mathcal{D}_\mathrm{train}} ; \tilde{\mathcal{D}}_\mathrm{test}, K \right) \text{ such that } \mathcal{C}_{\sim \mathcal{G}} \in \mathcal{D}_\mathrm{train}, \; \tilde{\mathcal{C}}_{\sim \tilde{\mathcal{G}}} \in \tilde{\mathcal{D}}_\mathrm{test}.
\end{equation}
Specifically, the new corpus $\tilde{\mathcal{C}}$ may result from two main scenarios with respect to an unseen new corpus and updates to a corpus:

\paragraph{Adaptation to new corpora} 
Adaptation to a new corpus refers to the trained IR models that may be hard to adapt to a corpus of new domains in the absence of supervised data \cite{thakur2beir}. 
The goal of improving generalizability on the adaptation to a new corpus for a neural IR model $f$ under top-$K$ ranked results can usually be formalized as:
\begin{equation} 
    \max \mathcal{R}_M\left(f_{\mathcal{D}_\mathrm{train}^{o}} ; \mathcal{D}_\mathrm{test}^{n}, K \right) \text{ such that } \mathcal{C}_{\sim \mathcal{G}}^{o} \in \mathcal{D}_\mathrm{train}^{o}, \; \mathcal{C}_{\sim \tilde{\mathcal{G}}}^{n} \in \mathcal{D}_\mathrm{test}^{n},
\end{equation}
where $o$ is the domain of the original corpus and $n$ is the new corpus domain.
Among the main solutions for adaptation to a new corpus are data augmentation \cite{wang2022gpl,izacard2021unsupervised,bonifacio2022inpars}, domain modeling \cite{yu2022coco}, architectural modifications \cite{kasela2024desire}, and scaling up the model capacity \cite{ni2022large}. 

\paragraph{Updates to a corpus} 
Updates to a corpus refer to the problem for the trained IR model to maintain its ranking performance under continuously arriving new documents \cite{chen2023continual}.
The goal of improving generalizability on the updates to a corpus for a neural IR model $f$ under top-$K$ ranked results can be formalized as:
\begin{equation} 
    \max \mathcal{R}_M\left(f_{\mathcal{D}_\mathrm{train}^{\sum^t_0}} ; \mathcal{D}_\mathrm{test}^{t+1} \right) \text{ such that } \mathcal{C}_{\sim \mathcal{G}}^{\sum^t_0} \in \mathcal{D}_\mathrm{train}^{\sum^t_0}, \; \mathcal{C}_{\sim \tilde{\mathcal{G}}}^{t+1} \in \mathcal{D}_\mathrm{test}^{t+1},
\end{equation}
where $t$ is the time session of each corpus update. 
The main solution for maintaining the ranking performance is continual learning, but the different paradigms of generative retrieval (GR) \cite{metzler2021rethinking} and dense retrieval (DR) \cite{guo2022semantic} lead to different solutions in the two settings. 

Below, we first introduce the evaluation metrics widely used for OOD generalizability on unseen documents. 
Then, we detail solutions for adaptation to a new corpus and updates to a corpus, respectively.

\subsubsection{Evaluation}
\label{OOD evaluation}

OOD generalizability of IR models on unseen documents is mainly measured by the ranking performance under the new corpus. 
For both adaptation to a new corpus and updates to a corpus, ranking performance is the common evaluation.
For updates to a corpus, previous work also evaluates the degree to which the old corpus is forgotten.

\paragraph*{Metrics for ranking performance} 
For adaptation to a new corpus and updates to a corpus, the ranking performance of IR models under unseen documents is evaluated by common metrics:
\begin{itemize}
    \item \textbf{NDCG} \cite{jarvelin2002cumulated} evaluates the quality of ranked results by measuring the gain of a document based on its position in the ranked list;
    \item \textbf{MRR} \cite{craswell2009mean} evaluates the performance of a ranking result by calculating the average of the reciprocal ranks of the first relevant document answer;
    \item \textbf{HIT} \cite{chen2023continual} evaluates the proportion of times a relevant document is found within a set of top-$N$ ranked results; and 
    \item \textbf{AP} \cite{JMLR:v24:22-0496} evaluates the average performance of the ranking performance metrics, overall new domains in adaptation to new corpus, and sessions in updates to a corpus; the ranking performance metric could be any of the above.
\end{itemize}

\paragraph{Metrics for the degree of forgetting the old corpus}
Updates to a corpus are an ongoing process with many sessions; they require that the model memorizes new data without forgetting the old. 
Therefore, some metrics have been proposed to evaluate the model performance from a time-series perspective.
\begin{itemize}
    \item \textbf{Training Time} \cite{cai2023l2r} evaluates the total time it takes for the IR model to learn new data while recalling old data;
    \item \textbf{Forget$_t$} \cite{cai2023l2r} evaluates how much the model forgets at session $t$:
    \begin{equation}
        \mathrm{Forget_t} =\frac{1}{t} \sum_{j=0}^{t-1} \max _{l \in\{0, \ldots, t-1\}}\left(p_{l, j}-p_{t, j}\right),
    \end{equation}
    where $p$ is the ranking performance under any common metrics; and 
    \item \textbf{FWT} \cite{cai2023l2r} evaluates how well the model transfers knowledge from one session to future sessions:
    \begin{equation}
        \mathrm{FWT} =\frac{\sum_{i=1}^{j-1} \sum_{j=2}^{T} p_{i, j}}{\frac{T(T-1)}{2}},
    \end{equation}
    where $T$ is the total number of sessions.
\end{itemize}

\subsubsection{Adaptation to new corpora}
Proposed solutions to the adaptation to a new corpus involve data augmentation, distributionally robust optimization, and domain-invariant projection. 

\paragraph{Data augmentation}
Data augmentation involves generating or modifying data in such a way that it bridges the gap between the source domain (the domain where the model was originally trained) and the target domain (the new domain where the model is to be applied). 
This can include techniques like synthesizing new data examples through transformations that maintain the integrity of the underlying patterns, translating examples from one domain to another, or creating semi-synthetic samples. 
The GPL \cite{wang2022gpl} uses an unsupervised domain adaptation method generative pseudo labeling, which combines a query generator with pseudo labeling from a cross-encoder.
HyperR \cite{cai2022hyper} performs a hyper-prompted training mechanism to enable uniform retrieval across tasks of different domains.

There is other work that conducts unsupervised pre-training by using large-scale positive and negative pairs with different data augmentation methods such as, query generation \cite{liang2020embedding, ma2021zero,chandradevan2024duqgen,thakur2beir,ma2021zero,sachan2023questions}, synthetic pre-training \cite{reddy2021towards,yu2022coco,izacard2021unsupervised,neelakantan2022text}, 
or synthetic relevance labels \cite{li2024domain,ge2023augmenting,ram2022learning, dai2022promptagator,liu2022challenges}.
Overall, data augmentation can enrich the training set to include more domain-relevant variations, thereby enhancing the model's ability to generalize across domains.

Very recently, LLMs for data augmentation have significantly enhanced IR models by enabling effective corpus adaptation \cite{laitz2024inranker,anand2023data,bonifacio2022inpars}.
\citet{Anaya-Isaza2022Data} explore various data augmentation strategies combined with transfer learning to improve MRI-based brain tumor detection accuracy.
\citet{Chen2023Cross-domain} develop a cross-domain augmentation network to enhance click-through rate prediction by transferring knowledge between domains with different input features.
\citet{Oh2023Data} propose a prompt-based data augmentation method using generative language models for creating synthetic parallel corpora, improving neural machine translation performance.

\paragraph{Domain modeling}
Domain modeling seeks to model the data from both the source and target domains into a common feature space where the differences between the domains are minimized. 
The idea is to learn a representation of the data that retains the essential information for the task at hand while discarding domain-specific features that might lead to bias or overfitting. 
By doing so, the model learns to focus on the underlying task without being distracted by differences between the domains. 
COCO-DR \cite{yu2022coco} use implicit distributionally
robust optimization to reweight samples from different source query clusters for improving model robustness over rare queries during fine-tuning.
Together with contrastive learning, this approach significantly improves the generalization of DRM over different corpora.
There have been many successive efforts to optimize for this problem, including MoDIR \cite{xin2022zero}, and ToTER \cite{kang2024improving}.    
\citet{xu2023berm} address the domain OOD challenge by modeling a single passage as multiple units with two objectives. 
One is the semantic balance between units and the other is the extractability of essential units.
Distributionally robust optimization helps in reducing the sensitivity of the model to changes in data distribution, thereby improving its adaptability.

Another way to deal with the new domain data is domain-invariant projection.
\citet{zhan2022disentangled} have been the first to use a relevance estimation module for modeling domain-invariant matching patterns and several domain adaption modules for modeling domain-specific features of multiple target corpora.
\citet{xian2023learning} propose a list-level alignment method, which aligns the distributions of the lists and preserves their list structure.
They also demonstrate the superiority of their method on theoretical grounds.
The domain-invariant feature space enables the model to perform well on the target domain using knowledge acquired from the source domain, thereby facilitating effective domain adaptation.

\paragraph{Architectural modifications}
By optimizing the architecture of an IR model, the model can be made to have good domain adaptability.
For instance, hybrid retrieval models have been employed to integrate out-of-domain semantics, enhancing zero-shot capabilities and using core strengths of foundational model features \cite{chen2022out,lee2023back}. 
Additionally, DESIRE-ME \cite{kasela2024desire} uses a mixture-of-experts to tailor retrieval strategies effectively across various domains.

Moreover, methods like employing search agents in hybrid environments or using knowledge distillation with hard negative sampling further support the development of IR systems that maintain high performance in unseen domains \cite{huebscher2022zero,formal2022distillation}. 
These strategies collectively enhance the adaptability and effectiveness of retrieval systems across a range of out-of-domain scenarios.

\paragraph{Scaling up the model capacity}
Scaling up the model capacity has been identified as a crucial approach that significantly boosts a model's ability to handle diverse data types and improve retrieval effectiveness across unfamiliar domains.

\citet{ni2022large} explore the impact of enlarging dual encoder architectures.
They demonstrate that larger models are not only more capable of handling complex queries but also exhibit enhanced generalization across different domains.
In a similar vein, \citet{lu2022ernie} introduce ``Ernie-search'' which uses a novel method of self on-the-fly distillation to bridge the gap between cross-encoder and dual-encoder architectures. 
This technique enhances the dual-encoder's performance by distilling knowledge from a more powerful cross-encoder, effectively scaling up the retrieval capacity without the direct computational cost typically associated with larger models.

\subsubsection{Updates to a corpus}
In this scenario, IR models need to be compatible with newly added documents, however, this can lead to catastrophic forgetting problems with old documents.
Therefore, continuous learning \cite{de2021continual} has become a dominant approach, which aims at adapting the model to the newly added data without losing the ability to understand the old data by quickly adapting to the new unlabeled (little labeled) data. 

\paragraph{Continual learning for generative retrieval}
In \acl{GR}, a sequence-to-sequence model is adopted to unify both the indexing and retrieval stages.
All document information is encoded into the model parameters.
The tight binding of the index to the retrieval module makes updating the index costly.
To tackle this challenge, DSI++ \cite{mehta2023dsi++} adapts a continual learning method for DSI \cite{tay2022transformer} to incrementally index new documents while maintaining the ability to answer user queries related to both previously and newly indexed documents.
After that, CorpusBrain++ \cite{guo2024corpusbrain++} uses a continual learning method on another \acl{GR} model called CorpusBrain \cite{chen2022corpusbrain}.
\citet{chen2023continual} propose CLEVER to incrementally index new documents while supporting the ability to query both newly encountered documents and previously learned documents.
CLEVER performs incremental product quantization \cite{jegou2010product} to update a partial quantization codebook, and use a memory-augmented learning mechanism to form meaningful connections between old and new documents.
Subsequent work on continuous learning has been devoted to the problem of guaranteeing to updates to a corpus for different \acl{GR} models, for example, DynamicIR \cite{yoon2023continually} and IncDSI \cite{kishore2023incdsi}.

\paragraph{Continual learning for dense retrieval}
In dense retrieval, the model needs to learn the representation space of the entire corpus and encode each document into an embedding to serve as the index.
Therefore, continual learning for dense retrieval should effectively adapt to the evolving distribution with the unlabeled new-coming documents, and avoid re-inferring all embeddings of old documents to efficiently update the index each time the model is updated.
L$^2$R \cite{cai2023l2r} uses backward-compatible representations to deal with this problem.
It first selects diverse support negatives for model training, and then uses a ranking alignment objective to ensure the backward-compatibility of representations.

\subsection{OOD Generalizability to Unseen Queries} 
\begin{figure}[t] 
  \centering 
    \resizebox{1.0\textwidth}{!}{%
    \begin{tikzpicture}[
      grow = right,
      level 1/.style={level distance=2cm, sibling distance=2.5cm}, 
      level 2/.style={level distance=2cm, sibling distance=1.2cm}, 
      level 3/.style={level distance=2cm, sibling distance=1.2cm}, 
      edge from parent path = {
        (\tikzparentnode.east) -| +(0.25,0) |- (\tikzchildnode.west)
      },
      bag/.style={
        text width = 8em,
        text centered,
        anchor = west,
        fill = white,
        draw = inkblue,
        thick,
        rounded corners,
        font=\small,
        minimum height = 2em},
      cite/.style={
        text width = 9em,
        text centered,
        anchor = west,
        fill = white,
        draw = gray,
        thick,
        rounded corners,
        font=\small,
        minimum height = 2em},
      ]

      \node[bag, fill = gray!20, draw = gray] {OOD generalizability on unforeseen corpus} 
          child  {
            node[bag] {Unseen query type}
              child {
                node [cite] {\cite{cohen2018cross,bigdeli2024learning,ma2021contrastive,lupart2023ms,sciavolino2021simple}}
              }
          }
          child  {
            node[bag] {Query variation}
              child {
                node [bag] {Hybrid methods}
                  child {
                    node [cite] {\cite{tasawong2023typo,campos2023noise,sidiropoulos2024improving,pan2023towards}}
                  }
              }
              child {
                node [bag] {Contrastive learning methods}
                  child {
                    node [cite] {\cite{zhuang2021dealing,liu2023mirs,ma2021contrastive}}
                  }
              }
              child {
                node [bag] {Self-teaching methods}
                  child {
                    node [cite] {\cite{chen2022towards,zhuang2022characterbert,zhuang2023typos}}
                  }
              }
          }

      ;
    \end{tikzpicture}%
    }
  \caption{Classification of OOD generalizability on unseen queries.} 
  \label{fig:query-tree} 
\end{figure}
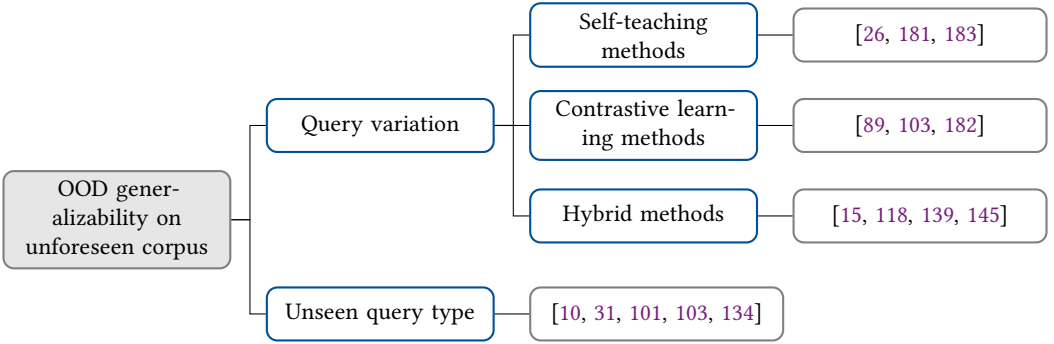

Deep-learning based models, constrained by their training data, often falter when faced with novel query formulations. 
Previous work has analyzed the generalizability of IR models under query variants and different query types \cite{wu2022neural,penha2022evaluating,zhuang2022robustness,liu2023robustness}, respectively. 
Next, we summarize prior work and how to improve the generalizability of IR models under unseen queries.
The specific methodology categorization is shown in Figure \ref{fig:query-tree}.

\subsubsection{Definition}
Generalizability to unseen queries indicates the capacity of an IR model to sustain its retrieval performance when confronted with new and unfamiliar query formulations. 
Enhancing a model's OOD generalizability with respect to unseen queries mainly involves improving the retrieval accuracy of the IR model across a variety of novel queries. 
Without loss of generality, given a test set $\tilde{\mathcal{D}}_\mathrm{test}$ comprising new queries $\tilde{\mathcal{Q}}$, which introduce a new distribution $\tilde{\mathcal{G}}$, the objective of augmenting the OOD generalizability of a neural IR model $f$ for unseen queries under top-$K$ ranked results can be formalized as:
\begin{equation} 
    \max \mathcal{R}_M\left(f_{\mathcal{D}_\mathrm{train}} ; \tilde{\mathcal{D}}_\mathrm{test}, K \right) \text{ such that } \mathcal{Q}_{\sim \mathcal{G}} \in \mathcal{D}_\mathrm{train}, \tilde{\mathcal{Q}}_{\sim \tilde{\mathcal{G}}} \in \tilde{\mathcal{D}}_\mathrm{test}.
\end{equation}
Specifically, the new queries $\tilde{\mathcal{Q}}$ may result from two main scenarios with respect to query variation and unseen query type:

\paragraph{Query variation}
Query variations refer to different expressions of the same information need.
The way in which information is expressed may impact the effectiveness of IR models \cite{penha2022evaluating}.
Some query variants, which introduce additional information, e.g., through query expansion, tend to enhance the retrieval performance \cite{belkin1995combining,benham2018towards}. 
Some introduce noise, such as typos, grammatical errors, and variations in word order, which often challenges the robustness of the IR model \cite{penha2022evaluating,campos2023noise,zhuang2021dealing}.
In this paper, we mainly focus on the latter one.

The goal of improving generalizability on the query variation for a neural IR model $f$ under top-$K$ ranked results can usually be formalized as:
\begin{equation} 
    \max \mathcal{R}_M\left(f_{\mathcal{D}_\mathrm{train}} ; G\left(\mathcal{D}_\mathrm{test}\right), K \right) \text{ such that }  \mathcal{Q}_{\sim \mathcal{G}} \in \mathcal{D}_\mathrm{train},  G\left(\mathcal{Q}\right)_{\sim \tilde{\mathcal{G}}} \in G\left(\mathcal{D}_\mathrm{test}\right),
\end{equation}
where $G(\cdot)$ is a query variation generator that can generate the query variant based on each query in $\mathcal{Q}$.
Among the main solutions for maintaining consistent performance when conformed with query variation are 
\begin{enumerate*}[label=(\roman*)]
\item self-teaching method, 
\item contrastive learning method, and
\item hybrid method. 
\end{enumerate*}

\paragraph{Unseen query type}
Unseen query type refers to unfamiliar query types with new query intents that have not been seen during model training.
The main solution for unseen query types is cross-domain regularization.
The goal of improving generalizability on an unseen query type for a neural IR model $f$ under top-$K$ ranked results can be formalized as:
\begin{equation} 
    \max \mathcal{R}_M\left(f_{\mathcal{D}_\mathrm{train}^{\tau_i}} ; \mathcal{D}_\mathrm{test}^{\tau_j}, K \right) \text{ such that } \mathcal{Q}_{\sim \mathcal{G}}^{\tau_i} \in \mathcal{D}_\mathrm{train}^{\tau_i}, \mathcal{Q}_{\sim \mathcal{G}}^{\tau_j} \in \mathcal{D}_\mathrm{test}^{\tau_j},
\end{equation}
where $\tau_i$ and $\tau_j$ are the different types of queries.

Next, we first introduce evaluation metrics that are widely used for OOD generalizability on unseen queries. 
Then, we detail existing solutions for query variation and unseen query type, respectively. 

\subsubsection{Evaluation}
OOD generalizability of IR models on unseen queries is mainly measured by the ranking performance under the new query set.
In addition to the metrics we mentioned in Section \ref{OOD evaluation} for measuring ranking performance, there are two other metrics for ranking. 
There are also specific metrics for unseen query types to evaluate differences in the performance of IR models across query types:

\paragraph{Metrics for ranking performance} 
In addition to MRR and NDCG, the ranking performance of the IR model under unseen queries is evaluated by other common metrics for query variation and unseen query type:
\begin{itemize}
\item \textbf{Recall} \cite{chen2022towards} measures the proportion of relevant documents that are successfully retrieved from the total amount of relevant documents available. 
\item \textbf{MAP} \cite{Larson_2010} quantifies the average precision of retrieval across different recall levels, effectively summarizing the precision at each point where a relevant document is retrieved.
\end{itemize}

\paragraph{Specific metrics for unseen query type} 
DR$_{OOD}$ evaluates the drop rate between the ranking performance on the original type of queries and the ranking performance on the unseen type of queries \cite{wu2022neural}:
\begin{equation} 
	DR_{OOD} = \frac{p_{OOD} - p_{IID}}{p_{IID}},
\end{equation} 
where $p_{IID}$ is the ranking performance on original type of queries and $p_{OOD}$ is the ranking performance on unseen type of queries.

\subsubsection{Query variation}
Solutions to the query variation challenge include 
\begin{enumerate*}[label=(\roman*)]
\item a self-teaching method, 
\item a contrastive learning method, and
\item a hybrid method. 
\end{enumerate*} 

\paragraph{Self-teaching methods}
The self-teaching approach to query variations focuses on distilling the matching capabilities of the IR model on the original clean query to the case of query variants.
These methods often align the model output for distillation.
\citet{chen2022towards} argue that the drift between query variations and original queries in model representation space affects the subsequent effectiveness of IR models.
Based on this, they propose RoDR, which calibrates the in-batch local ranking of query variants to that of the original query for the representation space alignment.
\citet{zhuang2022characterbert} also notice this issue and employ CharacterBERT \cite{el2020characterbert} as the backbone encoder to perform a character-level self-teaching method.
This method distills  knowledge from queries without typos into the queries with typos in a character embedding space.
ToRoDer \cite{zhuang2023typos} uses a pre-training method that uses bottlenecked information to recover the query variation.

\paragraph{Contrastive learning methods}
Contrastive learning-based approaches to query variation typically make the model robust to query variants by enhancing the supervision of query variants against the original query as well as positive and negative samples.
\citet{zhuang2021dealing} propose a simple typos-aware training method for BERT-based DRMs and NRMs.
During training, this method randomly selects query variants and migrated the supervised signals of the original queries directly for training.
By comparing the similarity between a query and its variations and other distinct queries with contrastive learning, \citet{sidiropoulos2022analysing} improve the robustness of IR models when encountering typos.
MIRS \cite{liu2023mirs} uses a robust contrastive method via injection of [MASK] tokens into query variations and encouraging the representation similarity between the original query and the variation.

\paragraph{Hybrid methods}
Some work enables IR models to perform better than previously when dealing with query variants through a combination of self-teaching and contrastive learning.
\citet{tasawong2023typo} propose a typo-robust representation learning method that combines contrastive learning with dual self-teaching achieving competitive performance.
CAPOT \cite{campos2023noise} introduces a notion of an anchoring loss between the unaltered model and the aligned model and designs a contrastive alignment post-training method to learn a robust model.
\citet{sidiropoulos2024improving} argue that previous work does not make full use of positive samples and employ contrastive learning with self-teaching that supports multiple positives.
There is also work that uses LLMs to enhance the ability to deal with typos \cite{pan2023towards}.

\subsubsection{Unseen query type}
Usually, part of the unseen query type is caused by unseen documents. Solutions to such problems are presented in listed in Section~\ref{sec: OOD corpus}.
In this subsection, we present additional solutions for unseen query types.

\citet{wu2022neural} analyze the robustness of existing ranking models in the face of unseen query types with five types of queries.
They find that most NRMs do not generalize well to unseen query types.
Even after training on multiple types of queries, NRMs still perform poorly when faced with a new kind of query.
Among all ranking models, traditional probabilistic ranking models, such as BM25 \cite{robertson1994some}, have the strongest generalizability to OOD query types, while NRMs are the worst.
\citet{liu2023robustness} analyze the robustness of the generative retrieval model and the dense retrieval model under different query types and find that both models are sensitive to query types.

To improve the generalizability to unseen query types for IR models, \citet{cohen2018cross} explore the use of adversarial learning as a regularization technique across different domains within the ranking task framework. By employing an adversarial discriminator and training a NRM across a limited number of domains, the discriminator acts to give negative feedback, thereby preventing the model from adopting domain-specific representations. 
\citet{bigdeli2024learning} propose to integrate two kinds of triplet loss functions into neural rankers such that they ensure that each query is moved along the embedding space, through the transformation of its embedding representation, in order to be placed close to its relevant document.
In this way, they provide the opportunity to jointly rank documents and difficult queries.
There has been some work focusing on the challenges of unseen query types in different scenarios \cite{ma2021contrastive,lupart2023ms,sciavolino2021simple}.

\section{Open Issues and Future Directions} 
\label{section Future}

In addition to the significant progres documented in Section~\ref{Sec: Adversarial Robustness} and~\ref{Sec: Out-of-distribution Robustness} above, robust neural IR presents several remaining challenges and opportunities for future research.

\subsection{Remaining Challenges and Issues}
In this subsection, we identify challenges and issues in robust neural IR with a special focus on adversarial robustness and OOD robustness, respectively. 

\subsubsection{Challenges on adversarial robustness in IR}
Many key issues in adversarial robustness in IR have not received much attention.

\paragraph{Penetration attacks against the whole ``retrieval-ranking'' pipeline}
As explained in Section~\ref{Sec: Adversarial Robustness} and~\ref{Sec: Out-of-distribution Robustness}, there is a considerable body of work that focuses on attacking the first-stage retrieval \cite{liu2023black,zhong2023poisoning} and re-rank stage \cite{liu2022order,wu2023prada} separately. 
Penetration attacks focus on exploiting vulnerabilities within the retrieval re-rank pipeline, a critical component in IR systems that ranks results based on relevance to the query. 
These attacks aim to manipulate rankings by identifying and exploiting weaknesses in the pipeline’s design or its underlying algorithms. 
This manipulation can result in irrelevant or malicious content being ranked higher than genuine content, compromising the integrity and reliability of the IR system. 

\paragraph{Universal attacks}
Universal attacks represent a form of adversarial threat that is particularly challenging due to its generalizability across different models and instances \cite{wallace2019universal}. 
Unlike targeted attacks that aim at specific vulnerabilities within a system, universal attacks exploit common weaknesses that are present across a wide range of systems. 
This makes it difficult to defend against them, as they require solutions that are not just effective for a single model or instance but across the entire spectrum of possible configurations. 
The development of robust defenses against universal attacks is therefore a significant challenge that demands innovative approaches and a deep understanding of the underlying vulnerabilities.

\paragraph{Dynamic attack scenarios}
Most prior work on adversarial attacks is based on static assumptions about search engines \cite{wu2023prada,liu2022order,chen2023towards}. 
However, search engines operate within a dynamic landscape, which may include changes such as the expansion or reduction of the corpus, and the demotion of documents suspected of spamming in their rankings. 
In search engines, the search engine results page (SERP) for a query is constantly changing. 
Research indicates that, in the dynamic environment of search engines, current attack methods struggle to maintain a consistent ranking advantage \cite{liu2023topic}. 
Therefore, designing attack methods that fully consider the dynamic nature of search engines is both practically significant and challenging. 
At the same time, it is worthwhile to explore the development of defense methods that evolve in tandem with updates to the search engine.

\paragraph{Gaming in search engines}
The phenomenon of ``gaming'' in search engines, where individuals or entities manipulate search results for competitive advantage, poses a significant challenge to maintaining the integrity and relevance of search outcomes \cite{kurland2022competitive}. 
This competitive manipulation not only undermines the quality of information presented to users but also erodes trust in the search engine's ability to deliver unbiased and accurate results. 
As search engines evolve into more sophisticated platforms, so too do the methods employed by those looking to exploit their algorithms for personal gain \cite{raifer2017information}. 
This ongoing battle between search engines and gamers necessitates the development of more advanced detection and mitigation techniques that can adapt to new gaming strategies, preserving the search engine's role as a reliable source of information.

\paragraph{Defense against unseen attacks}
In IR counter defenses, it is often the case that empirical defenses where the attack method is known, can yield good results \cite{liu2024perturbation}. 
Whereas in real scenarios, the attack methods are multiple and potentially unknown.
Defending against unseen attacks is a paramount challenge in enhancing adversarial robustness in IR systems. 
These attacks are particularly daunting because they exploit new or unknown vulnerabilities, making traditional defense mechanisms, which are often designed to combat known threats, ineffective. 
The key to overcoming this challenge lies in the development of adaptive, intelligent systems capable of anticipating potential threats and dynamically adjusting their defense mechanisms in real-time. 
Achieving this level of adaptability and foresight requires a profound shift in the current paradigms of security in IR, embracing more proactive and predictive approaches.

\paragraph{Defense in practice}
Implementing effective defense mechanisms in practice is a balancing act between effectiveness, efficiency, and cost. 
Effective defense strategies are those that can accurately detect and neutralize threats without significantly impacting the user experience or the relevance of search results. 
However, the computational resources required for these strategies often come with high costs and can affect the efficiency of the search engine, leading to slower response times and decreased user satisfaction. 
To address these challenges, search engines are increasingly turning to machine learning and artificial intelligence technologies that can provide scalable and cost-effective solutions \cite{leonhardt2024efficient,ma2022scattered}. 
These technologies enable the development of adaptive defense mechanisms that can learn from attack patterns and evolve over time, offering a dynamic approach to security that maintains the delicate balance between protecting the search engine and preserving its performance.

\subsubsection{Challenges on OOD robustness in IR}
Several key issues in OOD robustness in IR have not received much attention.

\paragraph{Reliance on large-scale data}
The reliance on large-scale data sets for training and evaluating IR systems poses significant challenges in ensuring OOD robustness. 
Large datasets often contain biases and do not necessarily represent the diversity of real-world scenarios \cite{kaplan2014big}, leading to models that perform well on seen data but poorly on unseen, OOD examples \cite{thakur2beir,yu2022coco}. 
Addressing this challenge requires innovative approaches to data collection and model training that prioritize diversity and real-world applicability, ensuring that IR systems remain reliable and effective across a broad range of OOD scenarios.

\paragraph{Lack of harmonized benchmarks for multiple OOD scenarios}
A major burden in enhancing OOD robustness in IR is the lack of harmonized benchmarks that accurately reflect the multitude of real-world, OOD scenarios. 
Without standardized benchmarks, it is difficult to assess the true robustness of IR systems across different contexts and to identify areas for improvement. 
Developing these benchmarks involves not only capturing a wide range of OOD scenarios but also ensuring that they are representative of the actual challenges faced by IR systems in practice. This effort is crucial for advancing the state of OOD robustness in IR.

\paragraph{Enhancing continuous corpus adaptation for DRMs}
As we have seen in Section~\ref{Sec: Adversarial Robustness} and~\ref{Sec: Out-of-distribution Robustness}, prior work mainly considers the problem of one-shot adaptation of DRMs to new corpora \cite{thakur2beir,cohen2018cross}. 
However, in real search engines, new documents are constantly added to the search engine and bring in a variety of new domains, which poses a challenge to the continuous adaptation ability of DRMs.
To address this, it is essential to develop DRMs that not only quickly adapt to new corpora in a one-shot learning scenario but also continuously learn and adjust as new data is introduced. 
This requires innovative approaches that can dynamically update the models in an incremental fashion without the need for frequent retraining from scratch. 
Techniques such as online learning, transfer learning, and meta-learning can play pivotal roles in enhancing the continuous corpus adaptation of DRMs, ensuring they remain effective and relevant in the ever-changing search landscape.

\paragraph{Improving OOD generalization of NRMs}
Approaches to improve OOD generalization ability mainly target neural retrieval models~\cite{thakur2beir,yu2022coco,zhan2022disentangled}. 
More generally, the OOD robustness of neural ranking models should be optimized along with retrieval models.
Enhancing the OOD generalization of NRMs involves developing models that can effectively handle queries that deviate significantly from the training distribution, thereby ensuring the retrieval of relevant and accurate results under a wide range of search scenarios. 
This challenge requires a multifaceted approach, incorporating advanced machine learning techniques such as robust representation learning, anomaly detection, and domain adaptation strategies. 
By prioritizing the OOD robustness of NRMs alongside DRMs, search engines can significantly improve their ability to serve high-quality, relevant content to users, even when faced with novel or unexpected queries.

\paragraph{Improving OOD robustness in practice}
Enhancing OOD robustness in practice is essential for maintaining the effectiveness, efficiency, and cost-effectiveness of search engines. 
To achieve this, search engines must be able to accurately identify and process queries that fall outside the typical distribution of observed data, ensuring that even uncommon or unseen queries return relevant and useful results. 
Implementing robust OOD handling mechanisms can significantly improve the quality of search results \cite{thakur2beir,liu2023robustness}, but this often requires sophisticated algorithms that can detect and adapt to OOD queries in real-time \cite{zhuang2022characterbert,penha2022evaluating}. 
While these algorithms can be computationally intensive, leading to higher operational costs, the investment in OOD robustness can ultimately enhance user satisfaction and trust in the search engine \cite{liu2023robustness,zhuang2023typos}. 
Moreover, optimizing these algorithms for efficiency can help mitigate additional costs, ensuring that improvements in OOD robustness also align with the search engine’s operational goals.

\subsection{Challenges and Opportunities Posed by Large Language Models}
LLMs have gained attention for their generative abilities.
There have been several works applying them to IR with good results. 
The introduction of LLMs may bring new robustness problems and also provide new solutions to known robustness problems.


\subsubsection{New challenges to IR robustness from LLMs}
Recently, there has been a lot of exploratory work on using LLMs for IR tasks. 
However, these attempts may pose new robustness challenges to LLMs-based IR methods due to the robustness issues of LLMs themselves.

\paragraph{New challenges to adversarial robustness}
When applied to IR systems, the adversarial vulnerability of the LLMs themselves is imported at the same time, as demonstrated by:

\begin{itemize}
\item \textbf{The vulnerability caused by hallucinations of LLMs.} 
There has been some work using the LLM as a ranking model \cite{sun2023chatgpt}. 
However, LLMs can generate plausible yet factually incorrect or irrelevant information, a phenomenon known as ``hallucination.'' 
Prior work has developed attacks against LLM-based ranking models, and vulnerability issues have been found with such models \cite{liu2024multi}.
In IR systems, such reliance can lead to the retrieval of misleading or inaccurate data, undermining the trustworthiness and reliability of the system. 
Addressing these hallucinations is crucial for maintaining the integrity of search results and the credibility of IR systems using LLMs \cite{shuster2021retrieval,wu2023ragtruth}.

\item \textbf{Defense costs associated with the scale and opacity of LLMs.} 
LLMs are not only large and complex but also often operate as black boxes with limited transparency into how decisions are made. 
This opacity complicates efforts to diagnose and mitigate vulnerabilities, making defensive measures both technically challenging and resource-intensive. 
As LLMs scale, the costs and complexity of implementing robust defensive strategies are likely to increase substantially.
\end{itemize}

\paragraph{New challenges to OOD robustness}
LLMs have shown biases and input sensitivities~\cite{ji2023survey,ferrara2023should}, and these will affect the OOD robustness of IR systems:

\begin{itemize}
\item \textbf{Bias in the corpus domain of LLMs.} 
The training process of LLMs may not comprehensively represent the diversity of real-world applications. 
This training leads to a bias towards the domain characteristics of the training data, which can degrade performance when the model encounters OOD queries or documents in IR tasks. 
Overcoming this bias is essential for the development of robust, generalizable IR systems with LLMs.

\item \textbf{Sensitivity of LLMs to query inputs.} 
LLMs can exhibit high sensitivity to slight variations in input \cite{chen-2024-sifo-arxiv,ji2023survey,ferrara2023should}, potentially leading to significantly different IR outcomes. 
This sensitivity can be particularly problematic in IR contexts where consistent and predictable retrieval performance is necessary. 
Enhancing the robustness of LLMs against input variation is a critical area for research and development.
\end{itemize}

\subsubsection{New opportunities for IR robustness via LLMs}
While the use of LLMs may introduce new robustness risks, the power of LLMs also provides new ideas for improving robustness. 
In the field of NLP, many works are used to enhance the robustness of NLP models using LLMs \cite{moraffah2024adversarial,helbling2023llm,zeng2024autodefense}, but not so much yet in IR.

\paragraph{New opportunities to adversarial robustness}
LLMs hold promise for improving the adversarial robustness of IR systems through their ability to generate and predict adversarial examples:

\begin{itemize}
\item \textbf{Generating adversarial examples with LLMs.} 
LLMs can be used to create sophisticated adversarial examples to train IR systems, thereby enhancing their ability to withstand malicious attacks.
By exposing systems to a wider array of adversarial tactics during training, LLMs can help develop more resilient IR models.

\item \textbf{IR model defense assisted with LLMs.} 
LLMs can assist in the development of defense mechanisms by predicting and countering adversarial strategies. 
Using LLMs in simulation environments to anticipate potential attacks allows for the proactive strengthening of IR systems.
\end{itemize}

\paragraph{New opportunities to OOD robustness}
The powerful generation and language understanding ability of LLMs can help to improve the OOD robustness of IR systems:

\begin{itemize}
\item \textbf{Synthesizing OOD training data with LLMs.} 
LLMs can generate diverse and complex datasets that mirror OOD scenarios \cite{bonifacio2022inpars,rajapakse-2023-improving}, providing IR systems with the training needed to better handle unfamiliar or novel situations. 
Such synthetic data can help improve the generalizability and robustness of IR models against OOD inputs.

\item \textbf{LLMs for OOD detection.}
Using LLMs' abilities in language understanding, they can be employed to detect and manage OOD queries effectively. 
By identifying queries that deviate from the training distribution, LLMs can trigger specialized handling for such cases, thereby enhancing the robustness and reliability of IR systems.
\end{itemize}

\section{Conclusion} \label{section Conclusion}
The landscape of IR has evolved significantly with the advent of neural methods. 
This evolution has brought with it new challenges in the robustness of IR systems. 
These robustness issues undermine the trust of users in search engines, making a critical concern for both researchers and practitioners in the field.

In this survey, we have explored the various forms of robustness in IR, focusing particularly on adversarial robustness and OOD robustness. 
We have also discussed the remaining challenges of these fields, as well as potential future directions for research.
This survey contributes to that journey by providing a structured overview of the current state-of-the-art, offering a roadmap for future research directions, and inspiring continued exploration and innovation in the field. 
While significant progress has been made in understanding and robustness of IR, there is still much work to be done. 
As the field continues to evolve, it will be crucial to develop robust defenses against these attacks, to ensure the integrity of search results and maintain the trust of users. 

In conclusion, robust neural IR represents a complex and multifaceted problem space, but also an opportunity for innovative research and development. 
As we look forward, the goal of developing IR systems that are not only robust but also adaptable, trustworthy, and user-centric is essential, promising to redefine the boundaries of what is possible in information retrieval.

\begin{acks}
This work was funded by the National Key Research and Development Program of China under Grants No. 2023YFA1011602 and 2021QY1701, 
the Strategic Priority Research Program of the CAS under Grants No. XDB0680102, 
the National Natural Science Foundation of China (NSFC) under Grants No. 62372431, 
the Youth Innovation Promotion Association CAS under Grants No. 2021100, 
the Lenovo-CAS Joint Lab Youth Scientist Project, 
and the project under Grants No. JCKY2022130C039.  
This work was also (partially) funded by 
the Hybrid Intelligence Center, a 10-year program funded by the Dutch Ministry of Education, Culture and Science through the Netherlands Organisation for Scientific Research, \url{https://hybrid-intelligence-centre.nl}, 
project LESSEN with project number NWA.1389.20.183 of the research program NWA ORC 2020/21, which is (partly) financed by the Dutch Research Council (NWO),
project ROBUST with project number KICH3.LTP.20.006, which is (partly) financed by the Dutch Research Council (NWO), DPG Media, RTL, and the Dutch Ministry of Economic Affairs and Climate Policy (EZK) under the program LTP KIC 2020-2023,
and
the FINDHR (Fairness and Intersectional Non-Discrimination in Human Recommendation) project that received funding from the European Union’s Horizon Europe research and innovation program under grant agreement No 101070212. 

All content represents the opinion of the authors,
which is not necessarily shared or endorsed by their respective employers and/or sponsors. 
\end{acks}

\appendix
\clearpage
\section{Source Selection}
\label{appendix:A}
We loosely followed the guidelines by \citet{kitchenham-2007-guidelines} for the selection of publications used in our survey.

\subsection{Sources}
We used the sources listed in Table~\ref{table:sources} for our survey.

\begin{table}[t]
\centering
\caption{Venues, journals, and repositories used for the survey.}
\label{table:sources}
\begin{tabular}{p{.75\textwidth}p{.2\textwidth}}
\toprule
\textbf{Source} & \textbf{Acronym or URL} \\
\midrule
 AAAI Conference on Artificial Intelligence  & AAAI \\
 Annual Meeting of the Association for Computational Linguistics & ACL \\
arXiv & \url{https://arxiv.org}\\
 ACM Conference on Computer and Communications Security & CCS \\
 ACM International Conference on Information and Knowledge Management& CIKM \\
 European Conference on Information Retrieval & ECIR \\
 Conference on Empirical Methods in Natural Language Processing & EMNLP \\
 Foundations and Trends in Information Retrieval & FnTIR \\
 International Conference on Learning Representations & ICLR \\
 International Conference on Data Mining & ICDM \\
 International Conference on Machine Learning & ICML \\
 International Conference on the Theory of Information Retrieval & ICTIR \\
 International Joint Conference on Artificial Intelligence & IJCAI \\
 Information Processing \& Management & IPM \\
 ACM SIGKDD Conference on Knowledge Discovery and Data Mining & KDD \\
Annual Conference of the North American Chapter of the Association for Computational Linguistics & NAACL \\
 Conference on Neural Information Processing Systems & NeurIPS \\
SIGIR Conference on Research and Development in Information Retrieval & SIGIR \\
Transactions of the Association for Computational Linguistics & TACL \\
 ACM Transactions on Information Retrieval & TOIS \\
 Text Retrieval Conference & TREC \\
 International World Wide Web Conference& WebConf \\
\bottomrule
\end{tabular}
\end{table}

\subsection{Inclusion criteria}
For papers in the sources listed in Table~\ref{table:sources} we used the following criteria to include them in our survey:
\begin{enumerate}[label=(IC\arabic*)]
    \item The paper proposes a definition of one or several robustness notions in the context of \ac{IR}.
    \item The paper proposes an approach to improving the robustness of an \ac{IR} model.
    \item The paper proposes a method to evaluate one or several robustness notions in the context of \ac{IR}.
    \item The paper presents one or several benchmarks for assessing the robustness of \ac{IR} models.
    \item The paper presents a study that investigates the foundations of robustness of \ac{IR} models.
\end{enumerate}

\subsection{Exclusion criteria}
For papers in the sources listed in Table~\ref{table:sources} we used the following criteria to exclude them in our survey:
\begin{enumerate}[label=(EC\arabic*)]
    \item The paper is not written in English.
    \item The paper is not in the date range of January 2012 to July 2024.
    \item An extended version of the paper has been published, which subsumes its contents.
\end{enumerate}

\begin{table}[h]
  \caption{Statistics of datasets in BestIR benchmark. \#Doc denotes the number of documents in corpus; \#Q$_{\mathrm{train}}$ denotes the number of queries available for training; \#Q$_{\mathrm{dev}}$ denotes the number of queries available for development; \#Q$_{\mathrm{eval}}$ denotes the number of queries available for evaluation.}
  \label{tab:Benchmark}
  \small
  \begin{tabular}{l p{2.5cm} l rrrr}
    \toprule
  \textbf{Robustness} & \textbf{Type} & \textbf{Dataset} & \textbf{ \#Doc} & \textbf{\#Q$_{\mathrm{train}}$} & \textbf{\#Q$_{\mathrm{dev}}$} & \textbf{\#Q$_{\mathrm{eval}}$} \\
  \midrule
    \multirow{12}{*}{\begin{tabular}{@{}l}Adversarial\\ robustness\end{tabular}} & \multirow{5}{*}{Basic datasets} &  MS MARCO document \cite{nguyen2016ms} & 3.2M & 370K & 5,193  & 5,793  \\
    & & MS MARCO passage \cite{nguyen2016ms} & 8.8M & 500K & 6,980  & 6,837  \\
    & & Clueweb09-B \cite{clarke2009overview} & 50M & 150 & --  & --  \\
    & & Natural Questions \cite{kwiatkowski2019natural} & 21M & 60K & 8.8K  & 3.6K  \\
    & & TriviaQA \cite{joshi2017triviaqa} & 21M & 60K & 8.8K & 11.3K \\
    \cmidrule(r){2-7}
    & \multirow{3}{*}{\begin{tabular}{@{}l}Expansion of\\ datasets\end{tabular}} & TREC DL19 \cite{Craswell_Mitra_Yilmaz_Campos_Voorhees_2020} & -- & -- & 43  & --  \\
    & & TREC DL20 \cite{craswell2021overview} & -- & -- & 54 & --  \\
    & & TREC MB14 \cite{lin2013overview} & -- & -- & 50  & --  \\
    \cmidrule(r){2-7}
    & \multirow{4}{*}{\begin{tabular}{@{}l}Off-the-shelf\\ datasets\end{tabular}} & ASRC \cite{raifer2017information} & 1,279 & -- & 31  & --  \\
    & & Q-MS MARCO \cite{liu2023topic} & -- & -- & 4,000  & -- \\
    & & Q-Clueweb09 \cite{liu2023topic} & -- & -- & 292  & -- \\
    & & DARA \cite{chen2023defense} & 164k & 50k &  3,490 & 3,489\\
    \midrule
  & \begin{tabular}[c]{@{}l@{}}Adaptation to a\\ new corpus\end{tabular} & BEIR \cite{thakur2beir} & \multicolumn{4}{c}{18 corpus}
  \\
    \cmidrule(r){2-7}
  \multirow{16}{*}{\begin{tabular}{@{}l}OOD\\ robustness\end{tabular}} & \multirow{4}{*}{\begin{tabular}{@{}l}Updates to a\\ corpus\end{tabular}} 
   & CDI-MS \cite{chen2023continual} & 3.2M & 370K & 5,193  & 5,793 \\
   & & CDI-NQ \cite{chen2023continual} & 8.8M & 500K & 6,980  & 6,837 \\
   & & LL-LoTTE \cite{cai2023l2r} & 5.5M & 16K & 8.5K   & 8.6K  \\
   & & LL-MultiCPR \cite{cai2023l2r} & 3.0M & 136K & 15K  & 15K \\
       \cmidrule(r){2-7}
       
   & \multirow{9}{*}{Query variation} & DL-Typo \cite{zhuang2022characterbert} & -- & -- & -- & 60 \\
   & & noisy-MS MARCO \cite{campos2023noise} & -- & -- & -- & 5.6k  \\
   & & rewrite-MS MARCO \cite{campos2023noise} & -- & -- & -- & 5.6k \\
   & & noisy-NQ \cite{campos2023noise} & -- & -- & -- & 2k  \\
   & & noisy-TQA \cite{campos2023noise} & -- & -- & -- & 3k  \\
   & & noisy-ORCAS \cite{campos2023noise} & -- & -- & -- & 20k  \\
   & & variations-ANTIQUE \cite{penha2022evaluating} & -- & -- & -- & 2k \\
   & & variations-TREC19 \cite{penha2022evaluating} & -- & -- & -- & 430 \\  
   & & \citet{zhuang2021dealing} & -- & -- & -- & 41k \\
    \cmidrule(r){2-7}
   & \multirow{2}{*}{Unseen query type} & MS MARCO \cite{nguyen2016ms} & -- & -- & -- & 15k  \\
   & & L4 \cite{surdeanu2008learning} & -- & -- & -- & 10k  \\
  \bottomrule
\end{tabular}
\end{table}
\section{The B\lowercase{est}IR Benchmark}
\label{appendix:B}

BestIR aims to provide a robustness evaluation benchmark for neural IR models.
In order to address the comprehensive robustness challenge, we construct benchmarks mainly in terms of seven types of two aspects of robustness, i.e., adversarial robustness and OOD robustness.

For adversarial robustness, the datasets are usually used both for adversarial defense and adversarial attack tasks.
There are three construction methodologies:
\begin{enumerate*}[label=(\roman*)]
\item Basic datasets, which are the original IR datasets that are directly performed attacks and defenses on; 
\item Expansion of datasets, which are extensions of the original dataset to model unknown queries or new documents; and
\item Off-the-shelf datasets, which are datasets customized for the task of adversarial attack or defense that can be used directly for evaluation.
\end{enumerate*}

For OOD robustness, datasets are used to evaluate the performance of the model under unseen documents and unseen queries, respectively.
For each scenario, there are two types of evaluation perspectives.
For unseen documents:
\begin{enumerate*}[label=(\roman*)]
\item Adaptation to a new corpus consists mainly of IR datasets from different domains; and
\item Updates to a corpus consist mainly of the same dataset sliced and diced based on factors such as time or randomly.
\end{enumerate*}
For unseen queries:
\begin{enumerate*}[label=(\roman*)]
\item Query variation consists of different variants of the same query intent, such as typos, changes in word order, and changes in the form of expression; and
\item Unseen query type consists mainly of the different types of queries in a dataset.
\end{enumerate*}

Table \ref{tab:Benchmark} summarizes the statistics of the datasets provided in BestIR.
BestIR is publicly available at \url{https://github.com/Davion-Liu/BestIR}.
There are lots of datasets available within each aspect for robustness challenges and continue growing.
We try to balance the assessment of each aspect of robustness to fully evaluate the model's abilities.
In the future, we will consider the issue of robustness in a broader sense and introduce datasets into BestIR.
Meanwhile, researchers can also focus on observing specific aspects of robustness performance according to this categorization according to their concerns.

\bibliographystyle{ACM-Reference-Format}
\bibliography{Reference}

\end{document}